\documentclass[aps,twocolumn,nofootinbib,groupedaddress,amsfonts,floatfix]{revtex4-1} 
\usepackage{graphicx,amsmath,amssymb,amstext,amsbsy,amsfonts,amsthm,color}
\pdfoutput=1

\usepackage{mathtools}
\usepackage{epsfig}
\usepackage{graphicx}
\usepackage{subfigure}
\usepackage{hyperref}
\usepackage{cancel}
\usepackage{balance}

\graphicspath{{Figures/}}

\usepackage{color}
\usepackage[dvipsnames]{xcolor}

\renewcommand{\eqref}[1]{Eqn.~(\ref{#1})}
\newcommand{\Beq}{\begin{eqnarray}}
\newcommand{\Eeq}{\end{eqnarray}}

\def\lsim{\mathrel {\vcenter {\baselineskip 0pt \kern 0pt \hbox{$<$} \kern 0pt \hbox{$\sim$} }}}

\def\gsim{\mathrel {\vcenter {\baselineskip 0pt \kern 0pt \hbox{$>$} \kern 0pt \hbox{$\sim$} }}}

\newcommand{\grchombo}{\mathtt{GRChombo}}

\newcommand{\dd}{\mathrm{d}}

\newcommand{\vp}{\varphi}  
\newcommand{\bs}{\boldsymbol} 
\newcommand{\pN}{{}_{\perp}N}
\newcommand{\rmi}{{\rm i}}

\begin{document}

\title{Local Continuity of Angular Momentum and Noether Charge for Matter in General Relativity}
\author{Robin Croft }
\email{rc634@cam.ac.uk}

\affiliation{\vspace{0.05cm} Centre for Theoretical Cosmology, Department
of Applied Mathematics and Theoretical Physics, University of Cambridge,
Wilberforce Road, Cambridge CB3 0WA, United Kingdom}

\begin{abstract}
Conservation laws have many applications in numerical relativity. However, it is not straightforward to define local conservation laws for general dynamic spacetimes due the lack of coordinate translation symmetries. In flat space, the rate of change of energy-momentum within a finite spacelike volume is equivalent to the flux integrated over the surface of this volume; for general spacetimes it is necessary to include a volume integral of a source term arising from spacetime curvature. In this work a study of continuity of matter in general relativity is extended to include angular momentum of matter and Noether currents associated with gauge symmetries. Expressions for the Noether charge and flux of complex scalar fields and complex Proca fields are found using this formalism. Expressions for the angular momentum density, flux and source are also derived which are then applied to a numerical relativity collision of boson stars in 3D with non-zero impact parameter as an illustration of the methods.

\end{abstract}

\maketitle

\subsubsection*{Conventions}

Throughout this work the metric has sign $\{-,+,+,+\}$ and physical quantities will be expressed as a dimensionless ratio of the Planck length $L_{pl}$, time $T_{pl}$ and mass $M_{pl}$ unless stated otherwise; for example Newtons equation of gravity would be written as
\begin{equation}
F=\frac{GMm}{r^2} \quad \rightarrow \quad\left(\frac{F}{F_{pl}}\right) = \frac{\left(\frac{M}{M_{pl}}\right)  \left(\frac{m}{M_{pl}}\right)}{\left(\frac{r}{L_{pl}}\right)^2},
\end{equation}
where $F_{pl} = M_{pl}L_{pl}T_{pl}^{-2}$ is the Planck force. Consequently $c$, $G$ and $\hbar$ take the numerical value of $1$. Additionally, tensor fields will be denoted using bold font for index free notation and normal font for the components. The dot product between two vector fields will be written interchangeably as $\bs{A}\cdot\bs{B}\leftrightarrow A^\mu B_\mu$ for readability. Additionally, $\nabla_\mu$ denotes the covariant derivative and $\partial_\mu$ is the partial derivative, both with respect to coordinate $x^\mu$. Finally, unless stated otherwise, Greek indices such as $\{\alpha, \beta, ..., \mu, \nu, ... \}$ label four dimensional tensor components whereas late Latin indices such as $\{i,j,k,...\}$ label three dimensional tensor components and early Latin indices such as $\{a,b,...\}$ label two dimensional ones.

\section{Introduction} \label{sect:intro}

Conservation laws play an important role in many areas of physics. For a general Lagrangian density $\mathcal{L}$, dependant on fields $\phi_i$ and derivatives $\partial_k \phi_i$ for $i \in \{1,2,...,m \}$ and $k \in \{1,2,...,n \}$, if a field transformation $\phi_i \rightarrow \phi_i + \delta \phi_i$ leaves the Lagrangian constant the Euler-Lagrange equations imply there is a conserved current $\boldsymbol{J}$, with zero divergence, given by
\begin{equation}
J^k = \sum_i \frac{\partial \mathcal{L}}{\partial (\partial_k \phi_i)}\delta \phi_i.
\end{equation}
In curved space a conserved current $\bs{J}$ satisfies $\nabla_\mu J^\mu = 0$. A charge $Q$ within 3-volume $V$ and a flux $F$ though $\partial V$, the boundary of $V$, can be associated with $\bs{J}$ as described later in Eqs.~(\ref{eq:Q_def_int}) and (\ref{eq:F_def_int}). If $\bs{J}$ is conserved then $Q$ is a conserved charge satisfying
\begin{equation}
\label{first_qf}\partial_t Q = {F}.
\end{equation}
This says the rate of change of a charge in a volume $V$ is equal to the flux across the boundary $\partial V$ of $V$. In the case that $\bs{J}$ has a non-zero divergence, $\nabla_\mu J^\mu \neq 0$, Eq.~(\ref{first_qf}) generalises to the continuity equation
\begin{equation}
\label{first_qfs}\partial_t {Q} = {F} - {S},
\end{equation}
where $S$ is defined in Eq.~(\ref{eq:S_def_int}); ${S}$ is the source of $\bs{J}$ in $V$ which can be understood as the destruction or creation of charge $Q$. Eq.~(\ref{first_qfs}) is a simplified version of Eq.~(\ref{qfs_system}), later referred to as the QFS system.

Evaluation of the continuity equations above, and their corresponding charges ${Q}$, have many uses in the study of fundamental fields in Numerical Relativity. One such use is the measurement of the Noether current $\bs{J}$ of a complex scalar/vector field which arises from a $U(1)$ gauge symmetry of the matter fields $\psi_j$ of the form $\psi_j \rightarrow \psi_j e^{\mathrm{i}a} \sim \psi_j + 
{\rm i} a \psi_j$ for some small constant $a$ and $j \in \{1,2,...,n\}$. The total charge $Q$, also called Noether charge in this case, is useful to track during numerical simulations as it gives insight into the numerical quality of a simulation. A violation of Noether charge conservation can arise from insufficient resolution in some region of the simulation or due to boundary conditions in a finite volume simulation. In the case of Sommerfeld (outgoing wave) boundary conditions \cite{Alcubierre:2002kk} we might expect charge to be transported out of a finite computational domain and the total charge $Q$ in the simulation should decrease. Monitoring only ${Q}$ within some volume $V$, it is impossible to tell whether Noether charge violation is due to a flux ${F}$ through the surface $\partial V$ or undesirable numerical inaccuracies such as dissipation. It is more useful to check whether the continuity Eq.~(\ref{first_qf}) (or equivalently Eq.~(\ref{first_qfs}) if there were a non-zero source term) is obeyed for a finite domain $V$; if this fails there is likely a problem as the continuity equations should be exactly observed for general spacetimes. 

Another use of the continuity equations is to measure the amount of energy-momentum belonging to matter fields within a volume $V$. This has many possible applications such as calculating the total energy or momentum of compact objects such as boson stars and neutron stars. The energy-momentum of matter obeys a conservation law in General Relativity as given by Penrose \cite{10.2307/2397365} where the considered spacetime is assumed to admit a Killing vector. In the case a Killing vector exists then a conserved current $\bs{J}$ associated with the energy-momentum tensor $\bs{T}$ can be identified. The current is $J^\mu = T^\mu_\nu \xi^\nu$ for some Killing vector $\bs{\xi}$ and satisfies $\nabla_\mu J^\mu = 0$. If $\bs{\xi}$ is a Killing vector then Eq.~(\ref{first_qf}) is the correct continuity equation and the charge $Q$ is conserved. In General Relativity the existence of Killing vectors is rare, reserved for spacetimes with special symmetries. Generic dynamic spacetimes with no symmetries, such as inspirals and grazing collisions of compact objects, have no easily identifiable Killing vector fields. If there is no Killing vector the divergence of $\bs{J}$ becomes $\nabla_\mu J^\mu = T^{\mu\nu}\nabla_\mu \xi_\nu$ and the source term ${S}$ is non-zero. Now Eq.~(\ref{first_qfs}) is the correct continuity equation and the charge $Q$ is no longer conserved. In section \ref{sect:mattercont} we will show how the choice of $\bs{\xi}$ affects the type of current $\bs{J}$, and therefore charge $Q$, obtained. While measures of energy or momentum are interesting in their own right, the measure of Eq.~(\ref{first_qfs}) within some volume ${V}$ can be a good measure of numerical quality of a simulation in a similar fashion to the measure of Noether charge mentioned already. 

When dealing with black hole spacetimes resolution requirements typically become very strict towards the singularity and lead to a local violation of Eqs.~(\ref{first_qf}) and (\ref{first_qfs}). This might not doom a simulation as for most physical applications in GR singularities are contained by an event horizon and are therefore causally disconnected from the rest of the simulation; a resolution problem in the vicinity of a singularity therefore may not propagate to the exterior. It could be helpful instead to consider a volume $\bar{V}$ equal to $V$ but removing a set of finite volumes $\tilde{V}_i$ which surround any singularities. Testing Eqs.~(\ref{first_qf}) and (\ref{first_qfs}) in volume $\bar{V}$ would then give a measure of the simulation resolution untainted by the resolution issues at a singularity.

Currently in Numerical Relativity it is common to measure energy-momentum in a localised region with with Eq.~(\ref{eq:Q_def_int}) for the charge $Q$ where the charge density $\mathcal{Q}$ is given in section \ref{sect:mattercont}. Examples of this can be seen in \cite{Sanchis_Gual_2019}, \cite{Di_Giovanni_2020},  \cite{PhysRevD.103.044059} and \cite{PhysRevD.96.024004}. While this is a good measure it neglects any radiation and the transfer of energy-momentum between matter and spacetime curvature; if the spacetime does not contain the corresponding Killing vector the charge cannot be treated as a conserved quantity. Instead the combination of variables $Q-S$, where $S$ is defined in Eq.~(\ref{eq:S_def_int}) and $\mathcal{S}$ in section \ref{sect:mattercont}, should be treated as a conserved quantity. Other popular methods to obtain the energy-momentum of a system include integrating asymptotic quantities such as the ADM mass and momentum, however these can not be used locally as they are defined in the limit of large radii only.

Recent work by Clough \cite{Clough_2021} evaluates $Q$, $F$ and $S$ for energy and linear momentum with the assumption that the approximate Killing vector $\bs{\xi}$ is a coordinate basis vector satisfying ${\partial_i \xi^j} =0$. Successful numerical tests of Eq.~(\ref{first_qfs}) are given for fixed and dynamic background simulations.

This paper builds on the work of \cite{Clough_2021} and generalises the system to measure angular momentum conservation and the conservation of Noether charges of complex scalar fields and spin-1 complex Proca fields. The assumption that the approximate Killing vector $\bs{\xi}$ is a basis vector satisfying $\partial_i \xi^j = 0$ is dropped and leads to a more general source term $\mathcal{S}$. The QFS system for angular momentum is also tested using fully non-linear numerical Relativity simulations of a spacetime consisting of two boson stars colliding in a grazing fashion.

This paper is organised as follows. In section \ref{sect:derive} the QFS system for a general non-conserved current is derived and section \ref{sect:sphere} explicitly expands the results for use with a spherical extraction surface. Even though no other extraction surfaces are considered, the results of \ref{sect:sphere} are easily adaptable  to other shapes. Section \ref{sect:noether} is a standalone derivation of the well known Noether charge density from the QFS perspective and goes on to find the flux variable; results for complex scalar fields and complex Proca fields are given. The application of the QFS system to energy momentum currents, angular momentum and energy are given in section \ref{sect:mattercont}. A fully non-linear test of the QFS system for angular momentum, using $\grchombo$ \cite{clough2015grchombo,Andrade2021} to perform Numerical Relativity simulations, is presented in section \ref{sect:results} along with a convergence analysis.

\section{Derivation of the QFS System} \label{sect:derive}
\begin{figure}[h]
{\includegraphics[width=1.0\columnwidth]{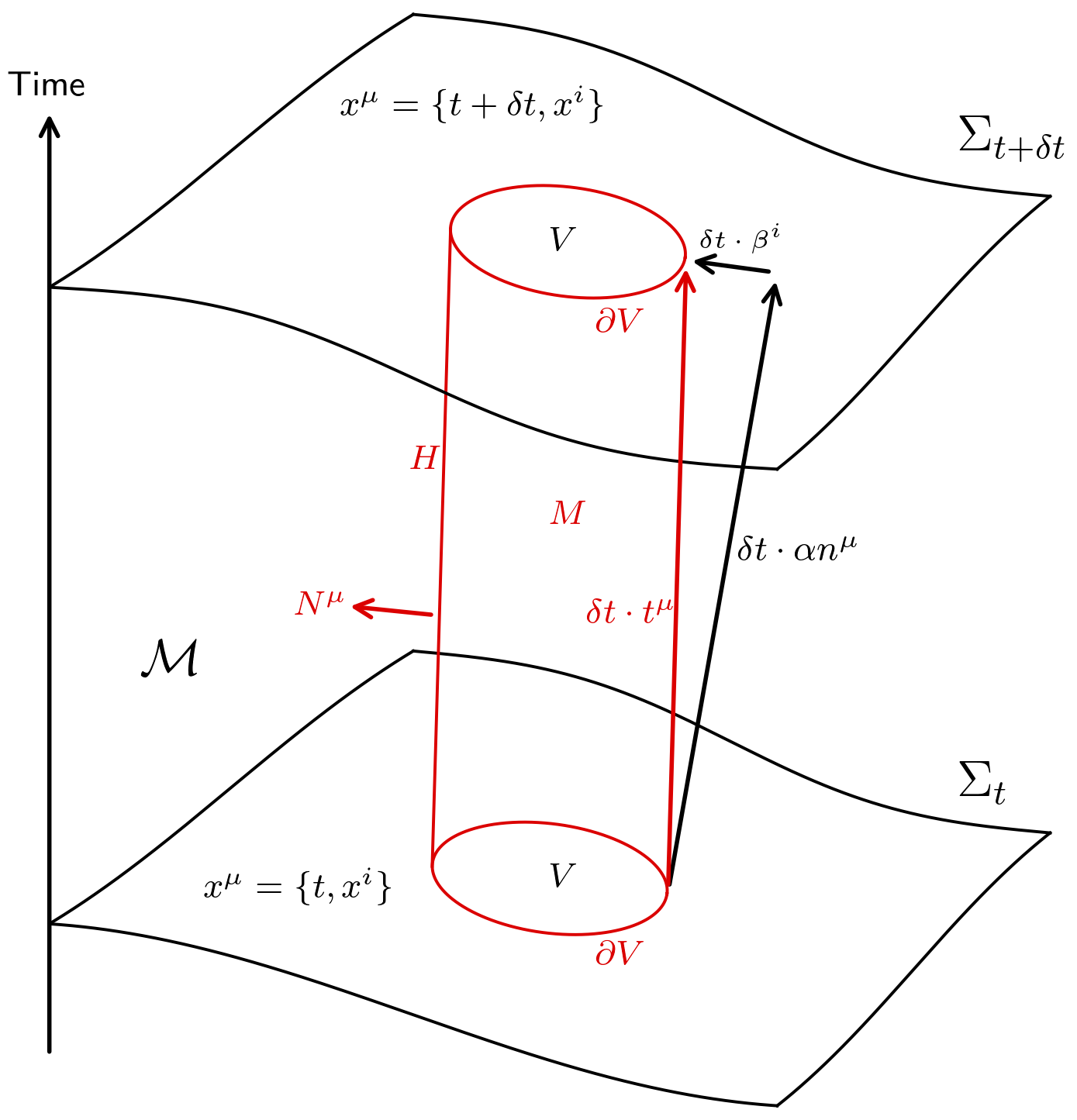}}
\caption{Diagram of relevant geometry for derivation of QFS system in section \ref{sect:derive} on manifold $\mathcal{M}$. $\Sigma_t$ is the spatial hypersurface at time $t$ and $\Sigma_{t+\delta t}$ is the the spatial hypersurface at a later time $t+\delta t$. $V$ is the coordinate volume, with surface $\partial V$, that we wish to use as an extraction volume on $\Sigma_t$. The sides of the red cylinder are $H$, defined by $\partial V$ evolved along integral curves of $\bs{t}=\bs{\partial_t}$. The interior of $H$ between times $t$ and $t+\delta t$ is $M$. Evolving $\partial V$ forward in time with $\bs n$, as demonstrated with the long black arrow, gives a different coordinate volume on $\Sigma_{t+\delta t}$ than on $\Sigma_t$.}
\label{fig:qfs_geometry}
\end{figure}

For a spacetime $(\mathcal{M},\bs{g})$ we start by defining a vector field $\bs{J}$ and subjecting it to the following continuity equation,
\begin{align}
\label{continuity_eqn_def}\nabla_\mu J^\mu = S,
\end{align}
where $S$ is a source term and describes the non-conservation of $\bs{J}$. In the case $S=0$ the current is conserved. We are interested in the charge density $\mathcal{Q}$ and source density $\mathcal{S}$ associated with $\bs{J}$ in a spatial 3-volume $V\in\Sigma$. Here $\Sigma$ is the usual 3-dimensional spacelike manifold $\Sigma$ consisting of the set of all points with constant time coordinate $t$, equipped with metric $\bs{\gamma}$. We are also interested in the flux density $\mathcal{F}$ through $\partial V$, the boundary of $V$ with metric $\bs{\sigma}$. $\Sigma$ is spanned by spatial coordinates $x^i$ related to the full spacetime coordinates $x^\mu$ by $ x^\mu = \{t,x^i \}$. The normal to $\Sigma$ is the unit co-vector $\bf{n}$ defined as, 
\begin{align}
\label{n_def}
n_\mu :&= \frac{\nabla_\mu t}{\sqrt{g^{\rho\sigma}\nabla_\rho t \nabla_\sigma t}} = -(\alpha,0,0,0), \\
\label{n_def2} n^\mu &= \frac{1}{\alpha}\left(t^\mu - \beta^\mu \right)= \frac{1}{\alpha} (1, -\beta^i ) , 
\end{align} 
where $(\alpha,\beta^i)$ are the usual lapse and shift from the ADM 3+1 spacetime decomposition \cite{2008}. The reader is directed to \cite{gourgoulhon20073+} for a comprehensive introduction to the $3+1$ decomposition. In Eq.~(\ref{n_def2}), $t^\mu = (1,0,0,0)$ is the future directed vector and is distinct from $n^\mu$. Time vector $\bs{t}$ is useful as its integral curves form lines of constant spatial coordinates. With this knowledge we can define the 4-volume $M$, the spatial 3-volume $V$ evolved along integral curves of $\bs{t}$ between times $t_0\leq t\leq t_0 + \delta t$ in the limit $\delta t\rightarrow 0$. Finally we define the 3-dimensional volume $H$, with metric $\bs h$. $H$ is the evolution of $\partial V$ along integral curves of $\bs{t}$ between times $t_0\leq t \leq t_0 + \delta t$ and is the 3-volume the flux crosses; clearly our definition of $H$ will affect our definition of flux density. There is no reason to choose the timelike vector $\bs{t}$, rather than $\bs{n}$, to evolve $V$ and $\partial V$ in time and both will result in a different definition of flux density. However, it is shown in appendix~\ref{sect:generality} that these two choices result in the same total integrated flux. A diagram summarising the relevant geometry can be found in Fig.~\ref{fig:qfs_geometry}.

With the relevant geometry discussed we can derive the QFS system, Eq.~(\ref{first_qfs}), by integrating Eq.~(\ref{continuity_eqn_def}) over $M$;
\begin{align} \label{master_integral}
\int_{M } \bs{\nabla}\cdot \bs{J} \sqrt{-g} \;\dd x^4 &= \int_{M } S \sqrt{-g} \;\dd x^4.
\end{align}
Let us start by using Gauss' theorem for curved space \cite{baumgarte_shapiro_2010} on the left hand side, 
\begin{align} \label{master_gauss}
\int_{M } \bs{\nabla}\cdot \bs{J} \sqrt{-g}\;\dd x^4  = \int_{\partial M } \hat{\bs{s}} \cdot \bs{J} \sqrt{{}^{(3)}{g}} \;\dd x^3 ,
\end{align}
with $\sqrt{{}^{(3)}g}$ being the volume element of a generic 3-surface and $\hat{\bs{s}}$ being the corresponding unit normal. Note that $\hat{\bs s}$ is outward directed when spacelike and inward directed when timelike. The integral of  $\bs{\nabla} \cdot \bs{J}$ over $M$ is now transformed to a surface integral of $\bs{J}$ over the compound 3-volume $\partial M$. This surface is split into an integral over $H$ and two integrals over $V$ at times $t_0$ and $t_0 + \delta t$. The integrals over $V$ give,
\begin{align}
&\left(\int^{(t=t_0)}_{V} - \int^{(t=t_0+\delta t)}_{V} \right)\bs{n} \cdot \bs{J} \sqrt{\gamma} \,\dd^3 x ,\nonumber \\
&=-\int_{V}\left[\left(\bs{n} \cdot \bs{J} \sqrt{\gamma}\right)_{t+\delta t} - \left(\bs{n} \cdot \bs{J} \sqrt{\gamma}\right)_{t}\right] \,\dd^3 x, \\
\label{q_int}&=-\delta t\,\partial_t \int_{V} \bs{n} \cdot \bs{J} \sqrt{\gamma}\,\dd^3 x ,
\end{align} 
where we made use of the fact that the coordinate volume $V$ is constant for all times due to it evolving in time with $t^\mu$. Here $\sqrt{\gamma}$ is the volume element on the spacelike manifold $\Sigma$. Now let us evaluate the integral over $H$, with metric $\bs{h}$ of signature $\{-,+,+\}$,
\begin{align} 
&\int_{H} \bs{N} \cdot \bs{J} \sqrt{-h} \,\dd^2 x\,\dd t, \nonumber \\
&=\int_{\partial V}  \bs{N} \cdot \bs{J} \sqrt{-h} \,\dd^2 x\int_t^{t+\delta t}\dd t, \\
\label{f_int}&=\delta_t \int_{\partial V} \bs{N} \cdot \bs{J} \sqrt{-h} \,\dd^2 x,
\end{align}
where $\bs{N}$ is the unit normal to $H$ as shown in Fig.~\ref{fig:qfs_geometry}. Given that $\bs{n}$ is not tangent to $H$ (but the time vector $\bs{t}$ is) we have $\bs{N} \cdot \bs{n} \neq 0$ and $\bs{N} \cdot \bs{t} =0$. This means we must normalise $\bs{N}$ with metric $\bs{g}$ and not $\bs{\gamma}$ as $\bs{N}$ is not tangent to $\Sigma$ and $\bs{g}(\bs{N},\bs{N})\neq \bs{\gamma}(\bs{N},\bs{N})$. In other words, $\bs{N}$ is a 4-vector in the case we time evolve $V$ with time vector $\bs{t}$ but $\bs{N}$ is a 3-vector if we time evolve with $\bs{n}$. Finally the right hand side source integral from (\ref{master_integral}) becomes,
\begin{align}
&\int_{M} S \sqrt{-g} \;\dd x^4 , \nonumber \\
&=\int_V S \sqrt{-g} \;\,\dd x^3 \int_t^{t+\delta t} \dd t,\\
 \label{s_int}&=\delta t \int_V S \alpha\sqrt{\gamma} \;\dd x^3 .
\end{align}
Combining Eqs.~(\ref{q_int}), (\ref{f_int}) and (\ref{s_int}) transforms Eq.~(\ref{master_integral}) into,
\begin{equation} \label{qfs_system}
\partial_t \int_{V} \mathcal{Q} \sqrt{\gamma} \,\dd^3 x  = \int_{\partial V} \mathcal{F} \sqrt{\sigma} \,\dd^2 x  
 - \int_{V} \mathcal{S} \sqrt{\gamma} \,\dd^3 x,
\end{equation}
where the density, flux density and source density ($\mathcal{Q}, \mathcal{F}, \mathcal{S}$) of angular momentum are defined as,
\begin{align}
\label{q_def}\mathcal{Q} :&= J^\mu n_\mu , \\
\label{flux_def}\mathcal{F} :&= \frac{\sqrt{-h}}{\sqrt{\sigma}}J^\mu N_\mu , \\
\label{source_def}\mathcal{S} :&= \alpha S. 
\end{align} 
The integrated version of these quantities can be written as
\begin{align}
\label{eq:Q_def_int}{Q} :&= \int_V \mathcal{Q}\sqrt{\gamma}\,\dd^3 x , \\
\label{eq:F_def_int}{F} :&= \int_{\partial V}  \mathcal{F}\sqrt{\sigma}\,\dd^2 x , \\
\label{eq:S_def_int}{S} :&= \int_V \mathcal{S}\sqrt{\gamma}\,\dd^3 x. 
\end{align} 
For later sections it is useful to split the normal vector $\bs{N}$ into its spacelike and timelike parts,
\begin{align}
\label{3plus1n}N_\mu = \pN_\mu - \bs{n} \cdot \bs{N} n_\mu,
\end{align}
where $\pN_\mu = \perp^\nu_\mu N_\nu$ is the projected part of $\bs{N}$ onto $\Sigma$ with $\bs{n}\cdot \bs{\pN}=0$ and $\bs{\perp}$ is the projection operator onto $\Sigma$,
\begin{align}
\label{perp} \perp^\nu_\mu = \delta^\nu_\mu + n^\mu n_\nu.
\end{align}
Using Eqs.~(\ref{3plus1n}) and (\ref{perp}) the flux term becomes, 
\begin{align}
\mathcal{F} &= \frac{\sqrt{-h}}{\sqrt{\sigma}}J^\mu (\pN_\mu - \bs{n} \cdot \bs{N} n_\mu), \\
  \label{3plus1f} &= \frac{\sqrt{-h}}{\sqrt{\sigma}} (\gamma^{\mu\nu} J_\mu N_\nu - \bs{n} \cdot \bs{N} \mathcal{Q}).
\end{align}
The term on the left arises from flux through the surface $\partial V$ and the term on the right is a consequence of the coordinate volume $V$ moving with respect to a normal observer with worldline traced by $\bs{n}$. Writing the flux term as above makes it obvious how the definition of flux depends on the 3-volume $H$ which determines $\sqrt{-h}$, $\sqrt{\sigma}$ and $\bs{N}$. Equivalently it can be seen that the density (\ref{q_def}) and source (\ref{source_def}) terms do not depend on the extraction surface.

\section{Application to Spherical extraction} \label{sect:sphere}
The numerical application of the QFS system in section \ref{sect:results} chooses a spherical volume $V$ to extract the angular momentum flux. Using standard Cartesian and spherical polar coordinates, $x^i_{\rm{cart}} = \{x,y,z\}$ and $x^i_{\rm{polar}}=\{r,\theta,\phi\}$ respectively, we can define $H$ as the coordinate volume $r=r_0$, $t_0\leq t \leq t+\delta t$. Thus, the normal $\bs{N}$ to $H$ is proportional to $\bs{\nabla}(r-r_0) = \bs{\nabla}(\sqrt{x^2 + y^2 + z^2}-r_0)$. Explicitly calculating the components $N_\mu$, and normalising to unity, with spherical polar spacelike coordinates gives,
\begin{align}
N_\mu &= \frac{\nabla_\mu r}{\sqrt{g^{\rho\sigma} \nabla_\rho r \nabla_\sigma r}}, \\
     \label{radial_N} &= \frac{1}{\sqrt{g^{rr}}}(0,1,0,0).
\end{align}
Note that if we had chosen $H$, the future evolution of $\partial V$, to be evolved along the unit vector $\bs{n}$ rather than time vector $\bs{t}$ then we would have obtained a different definition of $\bs{N}$ perpendicular to $\bs{n}$ rather than $\bs{t}$. The consequences of the alternate choice of $H$ are explored in appendix \ref{sect:generality}.

The density and source terms $\mathcal{Q}$ and $\mathcal{S}$ do not depend on the integration domain $V$, but the flux term $\mathcal{F}$ does. The calculation of the Flux term requires the evaluation of the volume element $\sqrt{-h}$ of $H$. Due to the choice that $H$ is the surface of constant radial coordinate, finding the metric of this surface is straightforward. Here we define spherical polar coordinates $x^\mu = \{t,r,\theta, \phi \}$ on $\mathcal{M}$ and $X^m = \{t,\theta, \phi \}$ spanning $H$. Projecting the 4-metric $\bs g$ onto $H$ we can write
\begin{equation}
\label{eqn:projmetric}{}^{(4)}h_{\mu\nu} = g_{\mu\nu}-N_\mu N_\nu,
\end{equation}
where ${}^{(4)}\bs{h}$ belongs to $\mathcal{M}$. The line element of a curve residing in $H$ can be equivalently evaluated in $\mathcal{M}$ or $H$; the pullback of ${}^{(4)}\bs{h}$ from $\mathcal{M}\vert_{r=r_0}$ to $H$ gives the 3-metric $\bs h$ belonging to $H$,
\begin{align}
\label{induced_metric}h_{mn}  &= {}^{(4)}h_{\mu\nu}\,\frac{\partial x^\mu}{\partial X^m} \frac{\partial x^\nu}{\partial X^n}, \\
 \label{eqn:hmn}&= \begin{pmatrix} g_{tt} & g_{t\theta} & g_{t\phi} \\ g_{\theta t} &  g_{\theta\theta}& g_{\theta\phi} \\ g_{\phi t} & g_{\phi\theta} & g_{\phi\phi} \end{pmatrix}.
\end{align}
A similar argument can be made for $\partial V$, the set of all points satisfying $r=r_0$ and $t=t_0$, with metric $\bs{\sigma}$. The metric components and volume element are
\begin{align}
\label{sigma metric}\sigma_{ab} &= \begin{pmatrix} g_{\theta\theta} & g_{\theta\phi} \\ g_{\phi\theta} & g_{\phi\phi} \end{pmatrix}, \\
\label{sigma det}\sqrt{\sigma}  &= \sqrt{g_{\theta\theta}g_{\phi\phi} - g_{\phi\theta}g_{\theta\phi}}\,.
\end{align}
Using Cramer's rule for the inverse of a matrix with Eqs.~(\ref{eqn:hmn}) and (\ref{sigma metric}) we get
\begin{align}
\label{eqn:htt}h^{tt} = \frac{{\sigma}}{h},
\end{align}
and reading from Eq.~(\ref{eqn:projmetric}) gives 
\begin{align}
{}^{(4)}h^{tt} &=      g^{tt} - N^t N^t, \\
\label{eqn:htt2}&= -\frac{1}{\alpha^2}\left(\frac{\gamma^{rr}}{g^{rr}} \right).
\end{align}
Similarly to Eq.~(\ref{induced_metric}), the pushforward of $\bs h$ on $H$ to ${}^{(4)}\bs h$ on $\mathcal{M}\vert_{r=r_0}$ gives
\begin{align}
{}^{(4)} h^{\mu\nu}  &= h^{mn}\,\frac{\partial x^\mu}{\partial X^m} \frac{\partial x^\nu}{\partial X^n}, \\
 &= \begin{pmatrix} h^{tt}&0&h^{t\theta }&h^{t\phi} \\ 0&0&0&0 \\ h^{\theta t}&0&h^{\theta \theta }&h^{\theta \phi} \\ h^{\phi t}&0&h^{\phi\theta }&h^{\phi\phi} \end{pmatrix},
\end{align}
which shows that $h^{tt}={}^{(4)}h^{tt}$. Combining Eqs.~(\ref{eqn:htt}) and (\ref{eqn:htt2}) it can be shown that
\begin{align}
     \label{rootminushexpand}\sqrt{-h} = \alpha \sqrt{\sigma} \sqrt{\frac{g^{rr}}{\gamma^{rr}}}.
\end{align}
Using this with Eq.~(\ref{radial_N}) we can expand Eq.~(\ref{flux_def}) for the flux term $\mathcal{F}$, 
\begin{align}
 \label{spherical_flux}\mathcal{F}&=  \frac{\sqrt{-h}}{\sqrt{\sigma}} J^\mu N_\mu = \frac{\alpha}{\sqrt{\gamma^{rr}}} J^r, 
 \end{align}
 but for practical purposes it is helpful to decompose this in terms of 3+1 variables as in Eq.~(\ref{3plus1f}),
 \begin{align}
   \mathcal{F} &= \alpha\frac{\sqrt{g^{rr}}}{\sqrt{\gamma^{rr}}} \left(\gamma^{\mu\nu}J_\nu {N}_\mu - \bs{n} \cdot \bs{N} \mathcal{Q}\right), \\
                   &=  \alpha\frac{\sqrt{g^{rr}}}{\sqrt{\gamma^{rr}}}\left(\gamma^{r\nu}J_\nu \frac{1}{\sqrt{g^{rr}}} +\alpha^{-1}\beta^r\frac{1}{\sqrt{g^{rr}}} \mathcal{Q}\right), \\
\label{3plus1fluxspherical}   &= \frac{1}{\sqrt{\gamma^{rr}}}\left(\alpha\gamma^{r\nu}J_\nu  + \beta^r \mathcal{Q}\right).
\end{align}
It is straightforward to re-derive the results of this section for other extraction volume shapes, such as cylinders or cubes/rectangles, with a redefinition of spacelike volume $V$ giving rise to different $\bs{N}$, $H$ and $\partial V$. It is wise to pick suitable coordinates adapted to the symmetry of the problem.

\section{Noether Currents} \label{sect:noether}

In this section we apply the previous results of the QFS system Eq.~(\ref{qfs_system})
to the continuity of Noether charge for both a complex scalar and the Proca field. The charge $\mathcal{Q}$ represents the number density of particles. Since the total particle number minus antiparticle number is always conserved the conservation law is exact and the source term $\mathcal{S}$ vanishes. 

Globally the total Noether charge should be conserved, however in numerical simulation this might not always be the case. Two common ways for non-conservation to occur are for the matter to interact with the simulation boundary conditions (often unproblematic) or some region of the simulation being insufficiently resolved (often problematic). Without knowledge of the Noether flux it is difficult to know what a change in Noether charge should be attributed to. If a large extraction volume containing the relevant physics shows a violation of Eq.~(\ref{qfs_system}) then the change in total Noether charge being due to boundary conditions can be ruled out and resolution is likely the culprit. 

When considering black hole spacetimes it is common for Noether charge to be dissipated as matter approaches the singularity; this is due to resolution requirements typically becoming very high in this region. The violation of Eq.~(\ref{qfs_system}) inside a black hole horizon might not cause any resolution problems for the black hole exterior however due to causal disconnection. If the extraction volume is modified to exclude finite regions containing any black hole singularities then Eq.~(\ref{qfs_system}) could be used to monitor the conservation of Noether charge away from troublesome singularities. This would be a good way of checking the resolution of a matter field in situations such as boson/Proca stars colliding with black holes or scalar/vector accretion onto a black hole.

\subsubsection{Complex Scalar Fields} \label{sect:noether1}
In this section we consider the conserved Noether current associated with a complex scalar field $\vp$ with Lagrangian
\begin{equation}
\label{csfl}\mathcal{L} = \left(\frac{1}{16 \pi }R -\frac{1}{2}g^{\mu\nu}\nabla_\mu \bar{\vp} \nabla_\nu \vp - \frac{1}{2}V(\vp \bar{\vp}) \right)\sqrt{-g},
\end{equation}
where $V$ is some real potential function. There is a U(1) symmetry where a complex rotation of the scalar field $\vp \rightarrow \vp e^{\rmi a}$, for constant $a$, leaves the action unchanged. The associated Noether current $\bs{J}$ can be found in \cite{liebling2017dynamical}, 
\begin{align}
J^\mu = \rmi g^{\mu\nu}(\vp\partial_\nu \bar\vp - \bar\vp\partial_\nu \vp),
\end{align}
and satisfies $\bs{\nabla}\cdot\bs{J} = 0$. The conservation is exact here which tells us the source term vanishes. In this case the Noether charge density (\ref{q_def}) and flux density (\ref{3plus1f}) are,
\begin{align}
\mathcal{Q}&= n_\mu J^\mu, \\
           &= \rmi (\vp n^\mu\partial_\mu \bar\vp - \bar\vp n^\mu\partial_\mu \vp), \\
           & = \rmi(\bar\vp \Pi - \bar\Pi \vp), \\
\mathcal{F} &= \frac{\sqrt{-h}}{\sqrt{\sigma}}(\rmi\gamma^{\nu \mu}N_\mu (\vp \partial_\nu \bar\vp - \bar\vp \partial_\nu \vp) - \bs{n} \cdot \bs{N}  \mathcal{Q}),
\end{align}
where $\Pi = -\bs{n} \cdot\bs{\nabla}\vp$ is the conjugate momentum of the scalar field. Using Eq.~(\ref{3plus1fluxspherical}) for a spherical extraction surface, and spherical polar spacelike coordinates $\{r,\theta,\phi\}$, this explicitly becomes
\begin{align}
\mathcal{F} &= \frac{1}{\sqrt{\gamma^{rr}}}(\rmi\alpha\gamma^{\nu r} (\vp \partial_\nu \bar\vp - \bar\vp \partial_\nu \vp) + {\beta^r} \mathcal{Q}).
       \end{align}

\subsubsection{Complex Vector Fields} \label{sect:noether2}
The Complex vector field $\bs{A}$, also called a Proca field, has Lagrangian
\begin{equation}
\mathcal{L} = \left(\frac{1}{16 \pi }R -\frac{1}{4}F^{\mu\nu}\bar{F}_{\mu\nu} - \frac{1}{2}V(A^\mu \bar{A}_\mu) \right)\sqrt{-g},
\end{equation}
where $F_{\mu\nu} = \nabla_\mu A_\nu - \nabla_\nu A_\mu$. Again $V$ is some real potential function. The action is invariant under a similar $U(1)$ complex rotation of the vector field $A^\mu \rightarrow A^\mu e^{\rmi a}$ for constant $a$. Following \cite{Minamitsuji_2018} this leads to the following Noether current $\bs{J}$,
\begin{align}
J_\mu = \rmi\left(\bar{A}^\nu F_{\mu\nu} - A^\nu \bar{F}_{\mu\nu} \right),
\end{align} 
which again satisfies $\bs{\nabla}\cdot \bs{J} =0$ and the source term vanishes. Defining a $3+1$ decomposition compatible with \cite{Zilh_o_2015} gives,
\begin{align}
A_\mu &:= n_\mu \Phi + a_\mu, \\
\Phi &= -A_\mu n^\mu, \\
a_\mu &= \perp^\nu_\mu A_\nu, \\
F_{\mu\nu} &:= n_\mu E_\nu - n_\nu E_\mu + B_{\mu\nu}, \\
E_\mu &= \perp^\nu_\mu F_{\nu\alpha}n^\alpha, \\
B_{\mu\nu} &= \perp^\alpha_\mu \perp^\beta_\nu F_{\alpha\beta} = D_\mu a_\nu - D_\nu a_\mu,
\end{align}
where $\phi$, $\bs{E}$, $\bs{a}$ and $\bs{B}$ all belong to $\Sigma$. Additionally $\bs{D}$ is the covariant 3-derivative of $\Sigma$. Note that $\bs{F}$ has no time-time component as $n_\mu n_\nu F^{\mu\nu} =0$ from the anti-symmetry of $\bs{F}$. Using these, the Noether charge (\ref{q_def}) becomes,
\begin{align}
\mathcal{Q}&= n_\mu J^\mu, \\
           &=\rmi\left( n^\mu \bar{A}^\nu F_{\mu\nu} - n^\mu A^\nu \bar{F}_{\mu\nu}  \right), \\
           &=\rmi\left( n^\mu \bar{a}^\nu n_\mu E_\nu - n^\mu a^\nu n_\mu \bar{E}_{\nu}  \right), \\
           &=\rmi\left( a^\nu \bar{E}_{\nu}  - \bar{a}^\nu E_\nu  \right).
\end{align}
 Using Eq.~(\ref{3plus1f}), the Noether flux is,
\begin{align}
\mathcal{F} &= \frac{\sqrt{-h}}{\sqrt{\sigma}}(\overrightarrow{N}^\mu j_\mu - \bs{n} \cdot \bs{N}  \mathcal{Q}), \\
 &= \frac{\sqrt{-h}}{\sqrt{\sigma}} ( \rmi\overrightarrow{N}^\mu (\bar{A}^\nu F_{\mu\nu} - A^\nu \bar{F}_{\mu\nu}) - \bs{n} \cdot \bs{N}  \mathcal{Q}).
 \end{align}
 Expanding $\overrightarrow{N}^\mu \bar{A}^\nu F_{\mu\nu}$ using the 3+1 split,
 \begin{align}
 \overrightarrow{N}^\mu \bar{A}^\nu F_{\mu\nu} &= \overrightarrow{N}^\mu \bar{a}^\nu B_{\mu\nu} - \overrightarrow{N}^\mu\bar{\Phi} n^\nu n_\nu E_\mu, \\
                            &= \gamma^{\mu\rho }{N}_\rho( \bar{a}^\nu B_{\mu\nu} +  \bar{\Phi}E_\mu),\\
                            &= \gamma^{\mu\rho }{N}_\rho( \bar{a}^\nu (\partial_\mu a_\nu - \partial_\nu a_\mu) +  \bar{\Phi}E_\mu),
 \end{align}
 where the Christoffel symbols from $D_\mu$ in $B_{\mu\nu}$ cancel out. Putting this into the expression for the Proca Noether flux we get,
 \begin{align}
  \begin{split}\mathcal{F}&= \frac{\sqrt{-h}}{\sqrt{\sigma}} \{ \rmi\gamma^{\mu\rho}{N}_\rho [\bar{\Phi} E_\mu  -\Phi  \bar{E}_{\mu} +\bar{a}^\nu (\partial_\mu a_\nu - \partial_\nu a_\mu) \\ &\quad\quad\quad\quad\quad-  a^\nu (\partial_\mu \bar{a}_\nu - \partial_\nu \bar{a}_\mu)] - \bs{n} \cdot \bs{N}  \mathcal{Q}\},\end{split} 
 \end{align}
 and equation (\ref{3plus1fluxspherical}) gives the flux term for a spherical extraction surface,
 \begin{align} \begin{split}
\mathcal{F}&= \frac{1}{\sqrt{\gamma^{rr}}} ( \rmi\alpha\gamma^{\mu r} ( \bar{\Phi} E_\mu - {\Phi} \bar{E}_\mu + \bar{a}^\nu (\partial_\mu a_\nu - \partial_\nu a_\mu)  \\&\quad\quad\quad\quad\quad- {a}^\nu (\partial_\mu \bar{a}_\nu - \partial_\nu \bar{a}_\mu) ) + \beta^r  \mathcal{Q}),
 \end{split}\end{align}
 where spherical polar spacelike coordinates $\{r,\theta,\phi\}$ used.

\section{Energy-Momentum Currents} \label{sect:mattercont}
To find the current associated with energy-momentum we consider a vector field $\bs{J}$ defined with respect to a second vector field $\bs{\xi}$ and the stress tensor $\bs{T}$ by
\begin{align}
\label{Killing current}J^\mu := T^\mu_\nu \xi^\nu. 
\end{align}
Calculating the divergence of this vector leads to the following continuity equation,
\begin{align}
\nabla_\mu J^\mu &= \underbrace{(\nabla_\mu T^\mu_\nu )}_{=0}\xi^\nu + T^\mu_\nu \nabla_\mu  \xi_\nu, \\
\label{divJ}\nabla_\mu J^\mu &= T^{\mu\nu} \nabla_{(\mu}  \xi_{\nu)},
\end{align}
where (\ref{divJ}) shows the divergence vanishes if $\bs{\xi}$ is a Killing vector of the spacetime; a vanishing divergence corresponds to a conserved current with a zero source term. For more general spacetimes where $\bs{\xi}$ is not Killing, the right hand side of (\ref{divJ}) leads to a non-zero source term accounting for the transfer of energy-momentum between matter and spacetime curvature \cite{Clough_2021}. The choice of $\bs{\xi}$ dictates the type of energy-momentum current retrieved; for instance $\bs{\xi} = \bs{\partial_t}$ will correspond to an energy current $J^\mu = T^\mu_\nu (\partial_t)^\nu = T^\mu_t$ and the spacelike choice $\bs{\xi} = \bs{\partial_i}$ gives a momentum current $J^\mu = T^\mu_\nu (\partial_i)^\nu=T^\mu_i$ corresponding to the coordinate $x^i$. For an account of energy and linear momentum continuity see \cite{Clough_2021}.

\subsubsection{Angular Momentum} \label{sect:angmom}
The numerical test of the QFS system (\ref{qfs_system}) in section \ref{sect:results} measures the conservation of angular momentum. To do this we choose $\bs{\xi} = \bs{\partial_\phi}$, used in Eq.~(\ref{Killing current}), which is the coordinate basis vector of some azimuthal coordinate $\phi$. The angular momentum current is
\begin{align}
\label{angmomcurrent} J^\mu = T^\mu_\nu (\partial_\phi)^\nu=T^\mu_\phi.
\end{align}
Any spacetime with azimuthal symmetry (e.g. the Kerr spacetime) will have a vanishing source term as $\bs{\partial_\phi}$ is a Killing vector. This includes numerical simulations of matter in a fixed background. The example simulation in section \ref{sect:results} is the fully nonlinear grazing collision of two boson stars and $\bs{\xi}$ is not a Killing vector for finite distances from the collision centre. In this case the source term is non-zero. Using the standard 3+1 decomposition of the stress tensor \cite{gourgoulhon20073+}, \cite{alcubierre2008introduction} and explicitly expanding the density term from Eq.~(\ref{q_def}) gives,
\begin{align}
 \label{q_def_angmom}\mathcal{Q} &= T^\mu_\nu n_\mu ({\partial_\phi})^\nu, \\
       &= (S^\mu_\nu + S^\mu n_\nu + S_\nu n^\mu + n_\nu n^\mu) n_\mu (\partial_\phi)^\nu, \\
       &=  S_\nu n^\mu  n_\mu (\partial_\phi)^\nu, \\
       \label{bigq}&= -S_\phi, \\
       \label{q_explicit}&= y S_x - x S_y,
\end{align}
where $x$ and $y$ are Cartesian coordinates related to spherical polar coordinates in the usual way. Combining Eqs.~(\ref{3plus1f}), (\ref{angmomcurrent}) and (\ref{bigq}) we can get the angular momentum flux through a spherical extraction surface,
\begin{align}
 \mathcal{F} &= \frac{\sqrt{-h}}{\sqrt{\sigma}} (\gamma^{\mu\nu} T_{\rho\mu} (\partial_\phi)^\rho N_\nu + \bs{n} \cdot \bs{N} S_\phi),\\
             &= \frac{\sqrt{-h}}{\sqrt{\sigma}} (\gamma^{\mu\nu} S_{\phi\mu} N_\nu + \bs{n} \cdot \bs{N} S_\phi).
\end{align}
Using Eq.~(\ref{3plus1fluxspherical}), for a spherical extraction surface, the flux term becomes,
\begin{align}
 \mathcal{F} &= \alpha\frac{\sqrt{g^{rr}}}{\sqrt{\gamma^{rr}}} (\gamma^{\mu r} S_{\phi\mu} N_r -\frac{\beta^r}{\alpha}N_r S_\phi), \\ 
 &= \alpha\frac{\sqrt{g^{rr}}}{\sqrt{\gamma^{rr}}} (\gamma^{\mu r} S_{\phi\mu} N_r -\frac{\beta^r}{\alpha}N_r S_\phi), \\ 
  \label{final_flux} &= \frac{1}{\sqrt{\gamma^{rr}}}\left(\alpha \gamma^{\mu r}S_{\phi\mu} -\beta^r S_\phi\right)
\end{align} 
in spherical polar coordinates. The explicit expansion of the source term $\mathcal{S}$ is left for the appendix \ref{sect:source}, but the result is given here,
\begin{align}\label{s_explicit_angmom} 
\begin{split}\mathcal{S} &= \alpha S^\mu_{\nu}{}^{(3)}\partial_\mu \xi^\nu + \alpha S^\mu_{\nu} {}^{(3)}\Gamma^\nu_{\,\,\,\mu \sigma} \xi^\sigma \\&\quad- S_\nu \beta^i \partial_i \xi^\nu  + S_\nu \xi^\mu \partial_\mu \beta^\nu - \rho \xi^\mu \partial_\mu \alpha.
\end{split}
\end{align}
As noted in appendix \ref{sect:source}, when choosing a coordinate system to evaluate $\mathcal{S}$, if $\bs{\xi}$ is a coordinate basis vector then the $\partial_i \xi^j$ terms vanish. 

It would be simple to re-derive these results for linear momentum, by using $\bs{\xi} = \bs{\partial_i}$ for momentum in the $x^i$ direction for example, where $x^i$ is some Cartesian spatial coordinate. Results for linear momentum can be found in \cite{Clough_2021}.

\subsubsection{Energy} \label{sect:energy}
A local conservation system can also be applied to energy with the choice of an approximate Killing vector $\bs{\xi}$, $\xi^\mu = (\partial_t)^\mu = t^\mu =(1,0,0,0)$, and energy current
\begin{align}
\label{energycurrent} J^\mu = T^\mu_\nu t^\nu = T^\mu_t.
\end{align}
 Using the standard 3+1 decomposition of the stress-energy tensor from \cite{gourgoulhon20073+} or \cite{alcubierre2008introduction} the energy density $\mathcal{Q}$ is,
\begin{align}
\mathcal{Q} &= T^\mu_\nu n_\mu \xi^\nu, \\
            &= T^\mu_\nu n_\mu (\alpha n^\nu+ \beta^\nu), \\
            &= \alpha \rho - S_\mu \beta^\mu,
\end{align}
from Eq.~(\ref{q_def}). Similarly, combining Eqs.~(\ref{flux_def}) and (\ref{3plus1n}), the energy flux is,
\begin{align}\mathcal{F} &= \frac{\sqrt{-h}}{\sqrt{\sigma}}T^\mu_\nu N_\mu \xi^\nu, \\
            &= \frac{\sqrt{-h}}{\sqrt{\sigma}}T^\mu_\nu (\pN_\mu - \bs{n}\cdot \bs{N} n_\mu) (\alpha n^\nu + \beta^\nu), \\
    \begin{split} &= \frac{\sqrt{-h}}{\sqrt{\sigma}}(-\bs{n}\cdot \bs{N}\alpha\rho -  \pN_\mu S^\mu \alpha \\&\quad\quad\quad+ \bs{n}\cdot \bs{N} S_\mu \beta^\mu + \pN^\mu \beta^\nu S_{\mu\nu}), \end{split}
\end{align} 
and Eqs.~(\ref{radial_N}) and (\ref{rootminushexpand}) can be used for a spherical extraction surface,
\begin{align} 
 \mathcal{F} = \frac{1}{\sqrt{\gamma^{rr}}}(\alpha \rho \beta^r - \alpha^2 S^r   +  \alpha S^r_\mu\beta^\mu - \beta^\mu S_\mu \beta^r ). 
\end{align} 
 The source term is omitted here as the expression derived in appendix \ref{sect:source} assumes that $\bs{\xi}$ is spacelike. For energy continuity a timelike approximate Killing vector $\bs{\xi}$ is used and leads to a different expression for the source term that can be found in \cite{Clough_2021} along with the above density $\mathcal{Q}$ and flux term $\mathcal{F}$.

\section{Numerical Application} \label{sect:results}
To numerically test the QFS system, given in Eq.~(\ref{qfs_system}), for angular momentum an example spacetime consisting of colliding boson stars is simulated in 3D using $\grchombo$ \cite{clough2015grchombo,Andrade2021}. $\grchombo$ is a modern, open source, Numerical Relativity code with fully Adaptive Mesh Refinement (AMR) using the Berger-Rigoutsos block-structured adaptive
mesh algorithm \cite{PhysRevD.67.104005}. The CCZ4 constraint damping formulation \cite{PhysRevD.67.104005,PhysRevD.85.064040} is used with the moving puncture gauge \cite{PhysRevLett.96.111101,PhysRevLett.96.111102}. Time integration is done with 4th order Runge-Kutta method of lines.

\subsubsection{Numerical Setup of Simulations}

The boson stars are composed of a complex scalar field $\vp$, minimally coupled to gravity with the Lagrangian given in Eq.~(\ref{csfl}). Boson stars are stable self-gravitating spherically symmetric solutions of the Einstein-Klein-Gordon system in curved space; for a detailed review see \cite{liebling2017dynamical}. In this work, the Klein-Gordon potential is chosen to be $V=m^2 \vp \bar{\vp}$, where $m$ is the mass of a bosonic particle, leading to so called {\it mini boson stars}. The Kaup limit for the maximum mass of a mini boson star can be found numerically as approximately 
\begin{equation} M_{\mathrm{Kaup}} \sim 0.633 ~\frac{\hbar c}{G m} = 0.633 ~M_{pl}^2 ~m^{-1}, \end{equation}
where the physical constants are included for completeness, but have numerical value $1$ in Planck units. Notably, the maximum mass of a mini boson star scales inversely with the boson particle mass $m$. 

The Lagrangian in Eq.~(\ref{csfl}) with potential $V=m^2 \vp \bar{\vp}$ is unchanged up to an overall constant under a rescaling of the boson mass like $m \rightarrow b m$, for some dimensionless constant $b$, while simultaneously rescaling $x^\mu \rightarrow b^{-1} x^\mu$ for coordinates with dimension length/time. Consequently a mini boson star solution, categorised by the central scalar field amplitude $\vp_c$, represents a one parameter family of solutions with ADM mass and radius inversely proportional to $m$. To keep the choice of $m$ arbitrary the coordinates used in the simulation are $m x^\mu$, which are exactly Planck units in the case $m=1$ (i.e. the Planck mass).

To measure the charge associated with angular momentum, the following angular momentum measures are considered,
\begin{align}
\label{eq:Q_def_num}Q &:= \int_V \mathcal{Q} \sqrt{\gamma}\,\dd^3 x, \\
\label{eq:F_def_num}F &:= \int_{\partial V} \mathcal{F}\sqrt{\gamma}\,\dd^2 x,\\
\label{eq:S_def_num}S &:= \int_V \mathcal{S} \sqrt{\gamma}\,\dd^3 x , \\
\tilde{Q} &:= Q(t=0) + \int_0^{t} F \,\dd t, \\
\delta Q_S &:= \int_0^{t}S\,\dd t, \\
\label{eq:Q_hat_def_num}\hat{Q}&:= Q + \delta Q_S, \\
\bar{Q}&:= \tilde{Q} - \delta Q_S,
\end{align}
where $\mathcal{Q}$, $\mathcal{F}$ and $\mathcal{S}$ are defined in Eqs.~(\ref{q_def_angmom}), (\ref{final_flux}) and (\ref{s_explicit_angmom}) respectively. $\hat{Q}$ is the angular momentum modified by $\delta Q_S$; this is equivalent to absorbing the source term into $Q$. $\tilde{Q}$ is the initial angular momentum modified by the time integrated total flux. Equation~(\ref{qfs_system}) implies $\hat{Q}=\tilde{Q}$ exactly, and we define the relative numerical error measure $e_1$ by
\begin{align}\label{eq:r1def}
e_1&:= \frac{\hat{Q}-\tilde{Q}}{\hat{Q}},
\end{align}
which converges to zero in the continuum limit. We can alternatively define a different relative error
\begin{align}\label{eq:r2def}
e_2&:= \frac{{Q}-\bar{Q}}{{Q}},
\end{align}
where the source term is not absorbed into $Q$. Again, Eq.~(\ref{qfs_system}) implies that $Q = \bar{Q}$, or $e_2=0$, in the continuum limit.

\begin{figure}[h!]
{\includegraphics[width=0.9\columnwidth]{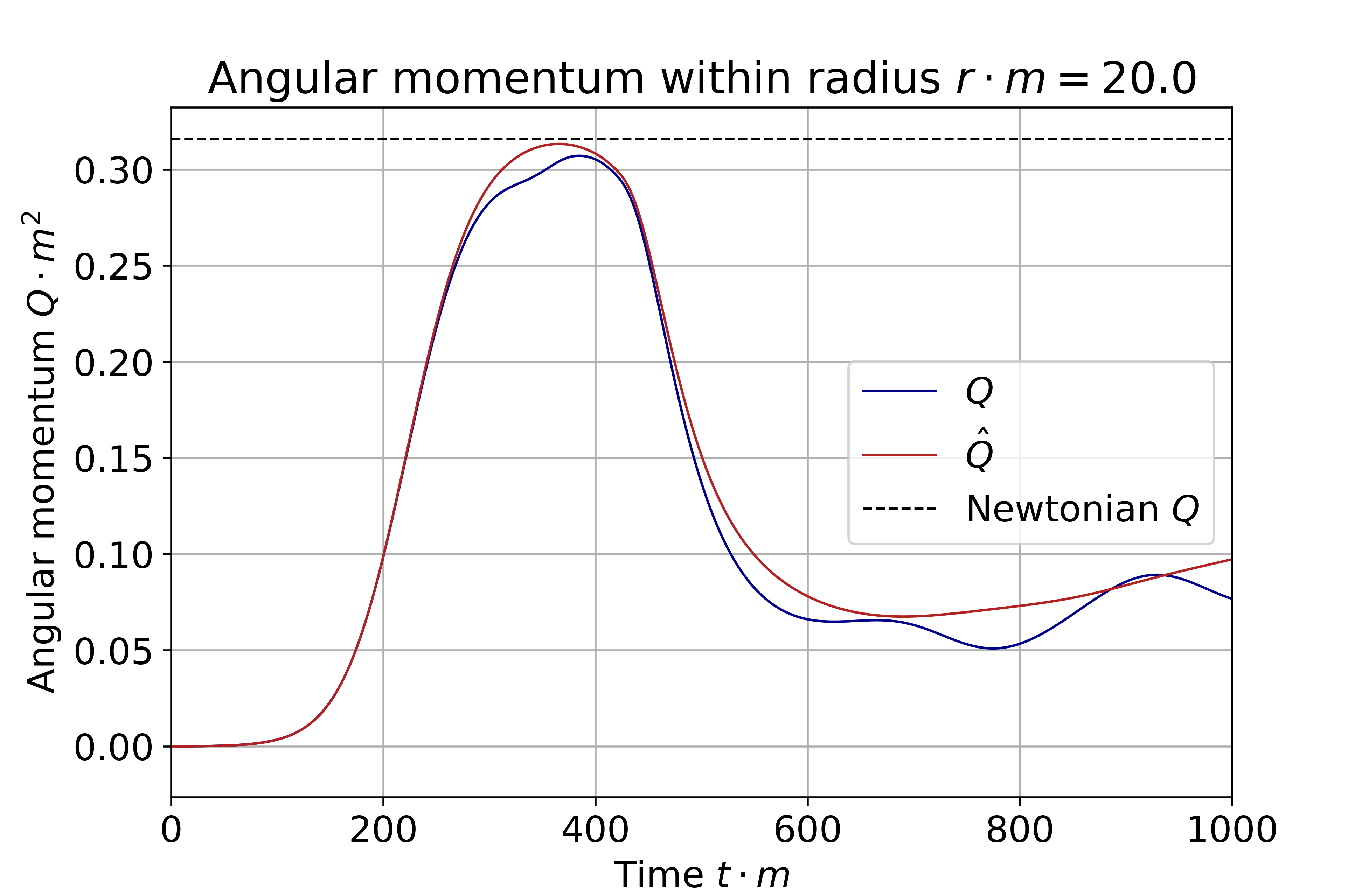}}
\caption{ Integrated angular momentum within radius $r<20 ~m^{-1}$. $Q$ is the  angular momentum integral in Eq.~(\ref{eq:Q_def_num}) and $\hat{Q}$ includes the source term as in Eq.~(\ref{eq:Q_hat_def_num}). The black dashed line indicates the Newtonian calculation for the angular momentum given in Eq.~(\ref{eq:newtQ}). The boson stars initially start outside the extraction radius. }
\label{fig:Q_20}
\end{figure}

\begin{figure}[h!]
{\includegraphics[width=0.9\columnwidth]{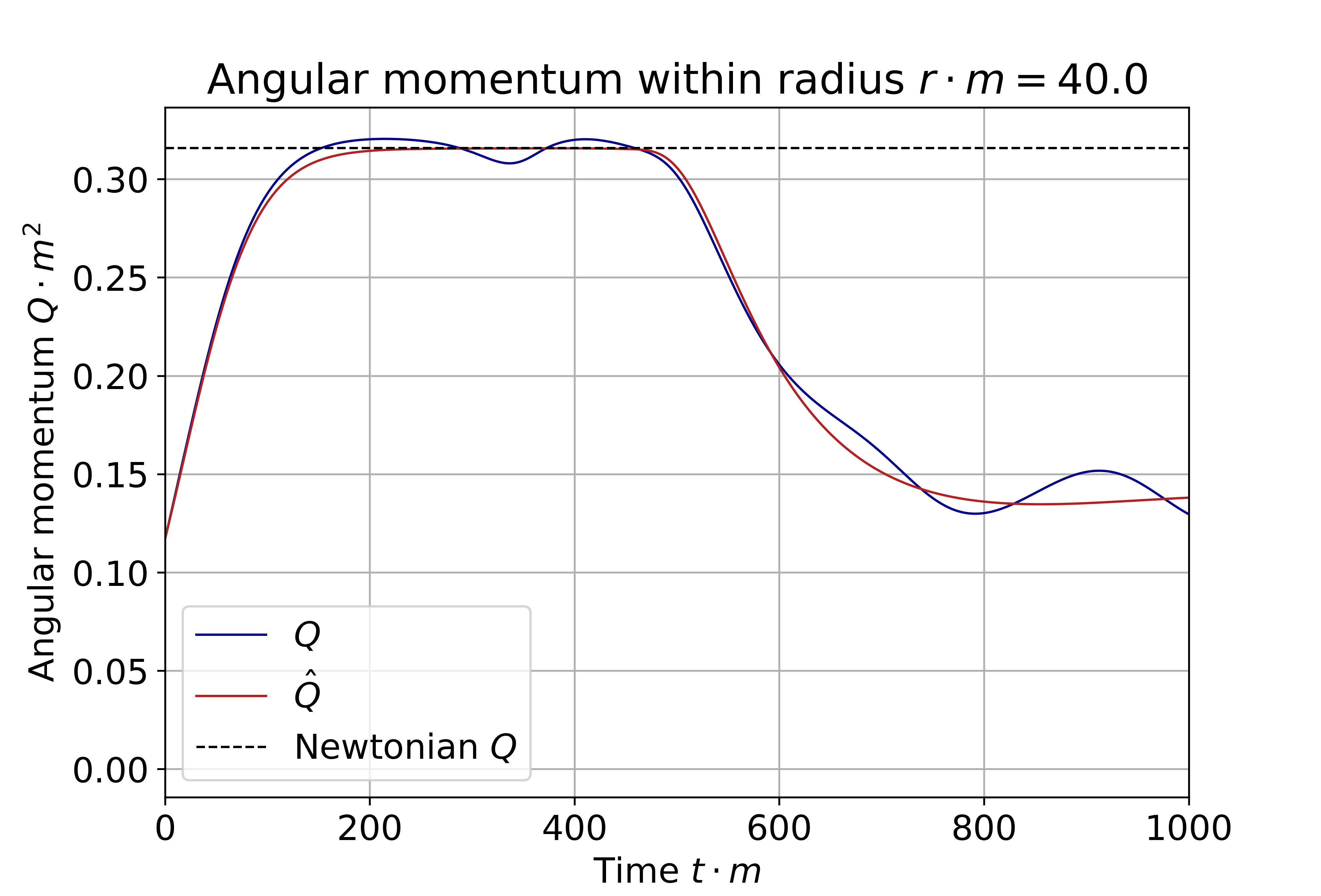}}
\caption{ Integrated angular momentum within radius $r<40 ~m^{-1}$. Quantities plotted are identical to Fig.~\ref{fig:Q_20}. The boson stars initially start intersecting the extraction radius.  }
\label{fig:Q_40}
\end{figure}

\begin{figure}[h!]
{\includegraphics[width=0.9\columnwidth]{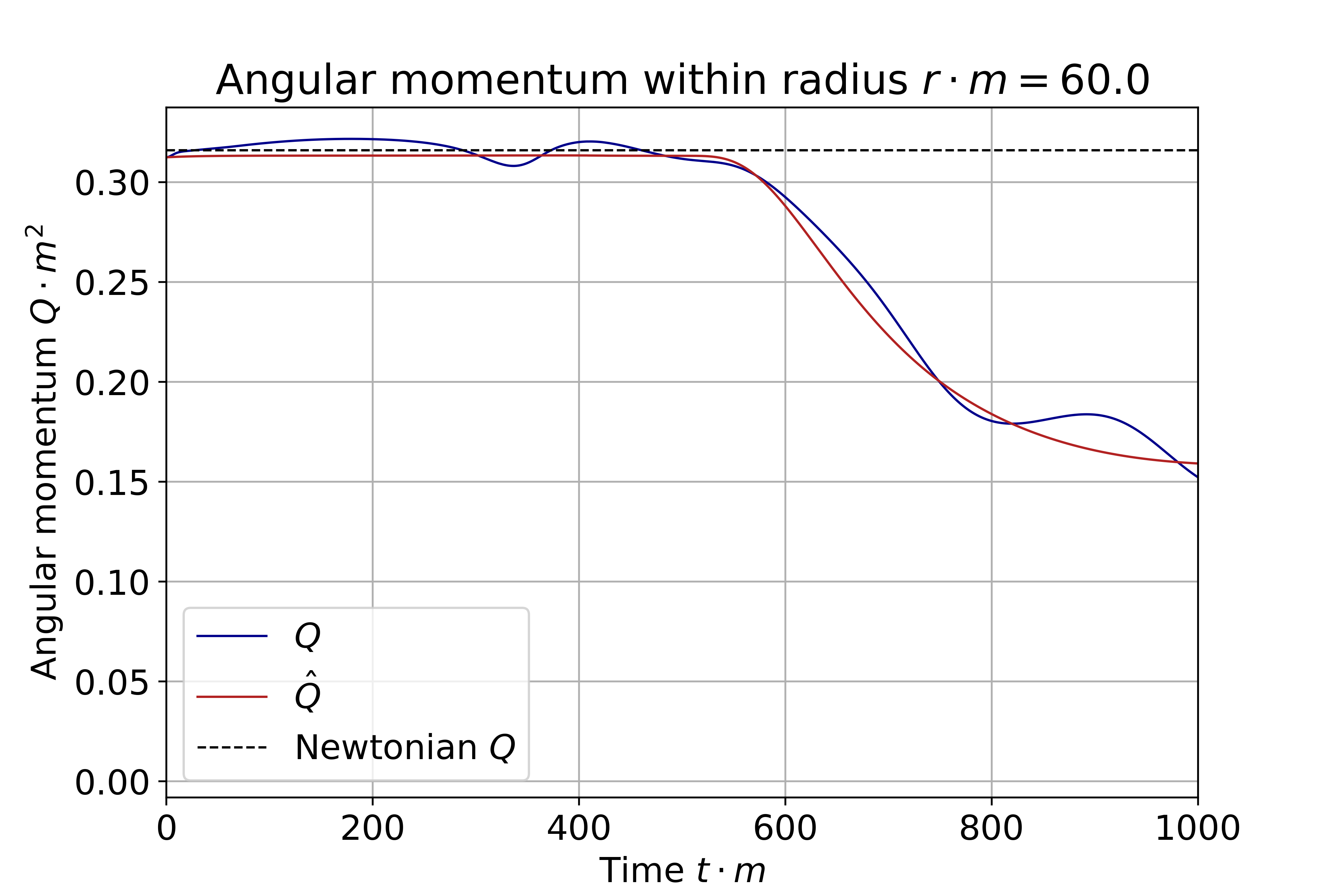}}
\caption{ Integrated angular momentum within radius $r<60 ~m^{-1}$. Quantities plotted are identical to Figs.~\ref{fig:Q_20} and \ref{fig:Q_40}. The boson stars initially start inside the extraction radius.  }
\label{fig:Q_60}
\end{figure}

The initial data of the numerical simulations consists of two boson stars, each with mass $M=0.395(0)~m^{-1}$, boosted towards each other in a grazing configuration. The data for two single boosted stars are superposed as in Ref.~\cite{helfer2021malaise} to minimise errors in the Hamiltonian and momentum constraints and spurious oscillations in the scalar field amplitudes of the stars. 
The physical domain is a cube of size $L=1024 ~m^{-1}$, the centre of this domain locates the origin of the Cartesian coordinates $x$, $y$ and $z$. The stars are placed at $x_0^i=\pm(40,4,0)~m^{-1}$ with respect to the centre of the physical domain, giving an initial impact parameter $d=8~m^{-1}$, and the boost velocity is $v^i=\mp(0.1,0,0)$ along the $x$ axis. The stars travel towards each other and undergo a grazing collision to form a short lived dense object at time $t \sim 375 ~m^{-1}$. Afterwards, much of the scalar field (and angular momentum) leaves the extraction radii as it is ejected to spatial infinity. Figures.~\ref{fig:Q_20}, \ref{fig:Q_40} and \ref{fig:Q_60} show the angular momentum within radii $r = \{20,40,60\}~m^{-1}$. The Newtonian angular momentum for this configuration is 
\begin{equation}\label{eq:newtQ}Mdv=0.316(0) m^{-2}\end{equation}
 which is in close agreement with $Q$ and $\hat{Q}$ in Figs.~\ref{fig:Q_20}, \ref{fig:Q_40} and \ref{fig:Q_60} while the matter is contained by the extraction radii. Given that we are dealing with a fully non-linear spacetime in general relativity there is no reason why the naive Newtonian angular momentum should agree so well with the numerically integrated values $Q$ or $\hat{Q}$; this could be due to the mass of the stars being $M=0.395(0) ~m^{-1}$, well below the Kaup limit $M_{\rm{Kaup}} \sim 0.633 ~m^{-1}$ and the mild boost velocities $v=0.1$. In the case that the star masses/densities and velocities tend to zero we expect general relativity to approach the Newtonian limit; conversely for large masses/densities and boost velocities the Newtonian estimate likely becomes less accurate. 

 Finally we note in Figs.~\ref{fig:Q_20}, \ref{fig:Q_40} and \ref{fig:Q_60} that the source-corrected density variable $\hat{Q}$ is less prone to oscillations than $Q$ and is closer to being constant at early times when no angular momentum flux is radiated. $\hat{Q}$ has another advantage over $Q$; at extraction radii sufficiently far from any matter $\hat{Q}$ will remain constant due to the flux $\mathcal{F}$ vanishing. $Q$ will only remain constant if the source term integral $\delta Q_S$ also remains constant which does not happen in general dynamic spacetimes, even for large extraction radii.

\subsubsection{Convergence Analysis}\label{sect:conv}

\begin{figure}[h]
{\includegraphics[width=0.95\columnwidth]{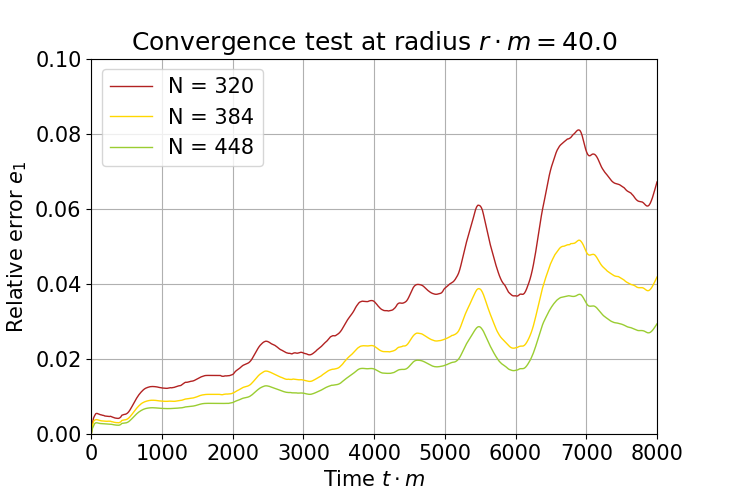}}
\caption{ Relative error $e_1$, from Eq.~(\ref{eq:r1def}), for the modified total angular momentum at extraction radius $r=40 ~m^{-1}$; the modified total angular momentum $\hat{Q}$ includes the source term. Figure includes four convergence simulations with $N\in\{320,384,448\}$ gridpoints along the coarse grid.}
\label{fig:r1}
\end{figure}
\begin{figure}[h]
{\includegraphics[width=0.95\columnwidth]{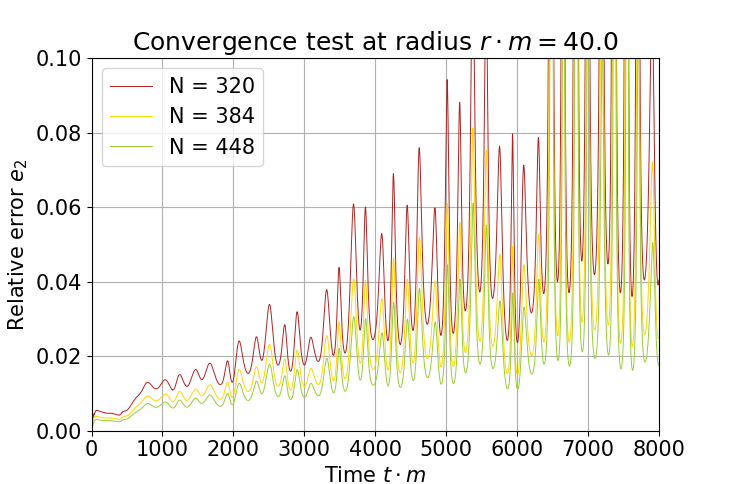}}
\caption{ Relative error $e_2$, from Eq.~(\ref{eq:r2def}), for the total angular momentum at extraction radius $r=40 ~m^{-1}$; the total angular momentum ${Q}$ excludes the source term. Figure includes four convergence simulations with $N\in\{320,384,448\}$ gridpoints along the coarse grid.} 
\label{fig:r2}
\end{figure}

Three numerical simulations are used to test the convergence of the angular momentum measures as the continuum limit is approached. They have $N\in\{320,384,448\}$ gridpoints on the coarsest level, named level $0$ with grid spacing $\Delta x_0 = L/N$. Each finer level, named level $n$, has grid spacing $\Delta x_n = 2^{-n} \Delta x_0$. Any gridpoints that fall inside radius $r= 200 ~m^{-1}$ are forced to be resolved by at least AMR level 1. Similarly, any points within radius $r<60 ~m^{-1}$ are resolved by at least AMR level 3; this modification quadruples the default resolution for $r<60 ~m^{-1}$ compared to level 1. These two radii have a $20\%$ extra buffer zone to ensure that AMR boundaries are outside and away from the desired radii. On top of this the AMR is triggered to regrid when a tagging criterion is exceeded; a description of the algorithm can be found in section 2.2.2 of \cite{Clough_2015}. The tagging criteria used in this paper involve gradients of the scalar field and spatial metric determinant; this loosely means as a region of spacetime becomes more curved, or matter becomes denser, the region is resolved with higher resolution. Figs.~\ref{fig:r1} and \ref{fig:r2} show the relative errors $e_1$ and $e_2$ for the convergence sequence; it can be seen that $e_1$, the relative error of $\hat Q$, is less prone to oscillations than $e_2$, the relative error of $Q$. The choice of enforcing AMR regridding to level 3 within $r<60 ~m^{-1}$ is very problem specific and the grid structure has been chosen carefully for the particular physical scenario to give higher resolution around the late time scalar field configuration at the origin; this enables accurate simulation of the extended object after merger. Simulations prior to this modification showed approximately five times higher relative error $e_1$ and much worse Noether charge conservation. The highest resolution simulation, with $N=448$, shows that the relative error $e_1$ is $3\%$ after $8000 ~m^{-1}$ time units.

\begin{figure}[h]
{\includegraphics[width=0.95\columnwidth]{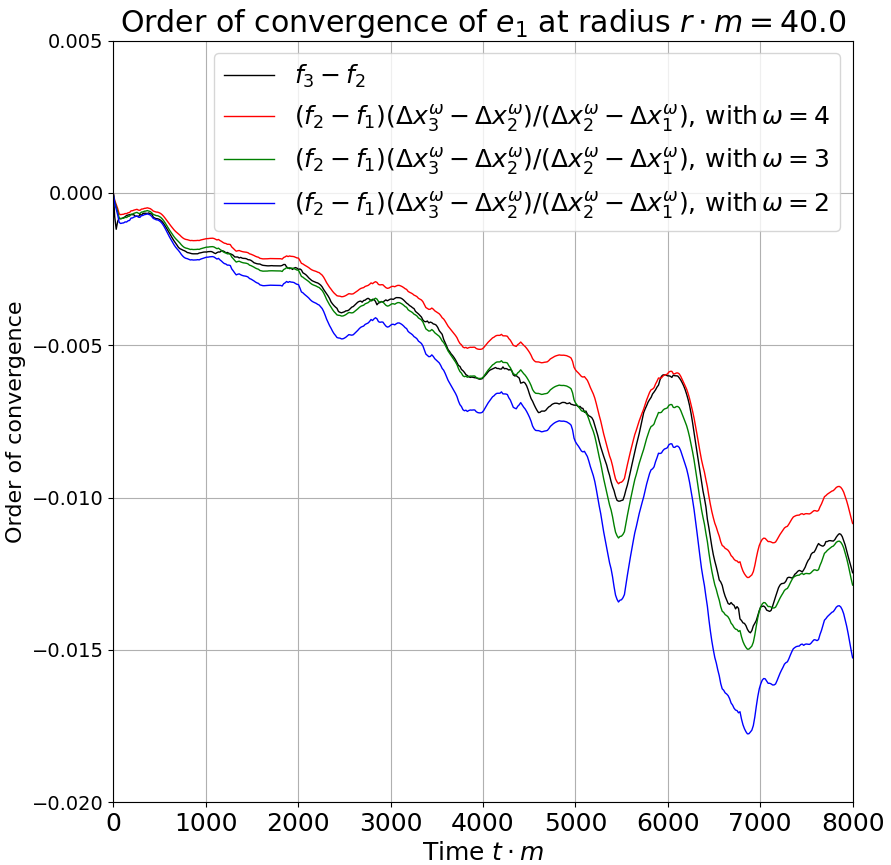}}
\caption{Estimating the order of convergence $\omega$ of the angular momentum error $e_1$ in Fig.~\ref{fig:r1} at extraction radius $r=40 ~m^{-1}$. The black curve shows the difference between $f_3$ and $f_2$; the relative error $e_1$ of the two highest resolution simulations in Section \ref{sect:conv}. The three coloured curves show the difference between the two lowest resolution simulations $f_2$ and $f_1$, but modified by $(\Delta x_3^\omega - \Delta x_2^\omega)/(\Delta x_2^\omega - \Delta x_1^\omega)$ in accordance with Eq.~(\ref{eq:trad_conv_def}), for three idealised orders of convergence $\omega=\{2,3,4\}$. The black curve is in best agreement with the green curve giving an estimate of $\omega=3$ for the order of convergence. }
\label{fig:234}
\end{figure}

We now obtain the order of convergence $\omega$ of $e_1$. It is convenient to express $e_1$ as three functions $\{f_1,f_2,f_3\}$ corresponding to the three different resolution simulations with $N=\{320,384,448\}$ and $f_\infty$ to denote the continuum limit solution. A traditional convergence analysis, as in \cite{PresTeukVettFlan92}, assumes that the numerical error of a function (i.e. difference from $f_\infty$) is dominated by a term proportional to $\Delta x_i^\omega$ for an order of convergence $\omega$; thus we can write
\begin{equation}\label{eq:convdef}f_i + E (\Delta x_i)^\omega = f_\infty\end{equation} 
for some constant coefficient $E$ for all resolutions $i$. Equation (\ref{eq:convdef}) with $i=\{1,2,3\}$ can be used to eliminate both $E$ and $f_\infty$ giving the well known result
\begin{equation}
\label{eq:trad_conv_def}\frac{f_3-f_2}{f_2-f_1} = \frac{ \Delta x_3^\omega-\Delta x_2^\omega }{ \Delta x_2^\omega-\Delta x_1^\omega }
\end{equation}
for ideal convergence. Figure~\ref{fig:234} shows $f_3-f_2$ and $(f_2-f_1 )(\Delta x_3^\omega - \Delta x_2^\omega)/(\Delta x_2^\omega - \Delta x_1^\omega)$ for three orders of convergence $\omega=\{2,3,4\}$; the two expressions should be equal for an ideal order of convergence $\omega$. It can be seen by eye that $\omega=3$ is the best estimate.

\begin{figure}[h]
{\includegraphics[width=0.95\columnwidth]{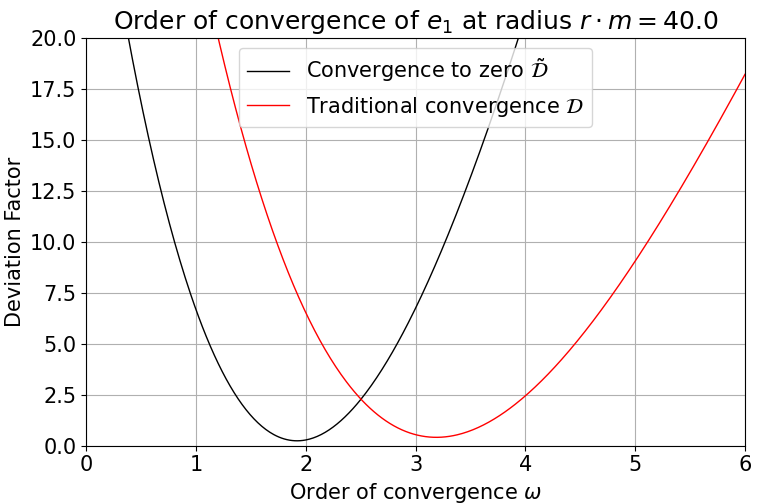}}
\caption{Estimating the order of convergence $\omega$ of the angular momentum error $e_1$ in Fig.~\ref{fig:r1} at extraction radius $r=40 ~m^{-1}$. The red curve gives the deviation from ideal traditional convergence $\mathcal{D}$, defined by Eq. (\ref{eq:tradconv}), as a function of $\omega$. The black curve shows the deviation from ideal convergence to zero $\tilde{\mathcal{D}}$, using the definition given in Eq. (\ref{eq:zeroconv}), as a function of $\omega$.  For both curves the estimated order of convergence is found by minimisation with respect to $\omega$; this gives $\omega=3.2$ for traditional convergence and $\omega=1.9$ for convergence to zero. }
\label{fig:D}
\end{figure}

To quantify the order of convergence, rather than guessing, we define the deviation factor $\mathcal{D}$ as
\begin{align}\label{eq:tradconv}
\mathcal{D}(\omega) = \int_{t_0}^{t_1}\left(\frac{f_3-f_2}{f_2-f_1} - \frac{ \Delta x_3^\omega-\Delta x_2^\omega }{ \Delta x_2^\omega-\Delta x_1^\omega }\right)^2\,\dd t,
\end{align}
which averages the violation of Eq.~(\ref{eq:trad_conv_def}) between times $t_0\leq t\leq t_1$. Figure \ref{fig:D} plots $\mathcal{D}$ versus $\omega$ with a red curve and the order of convergence can be estimated by minimising $\mathcal{D}(\omega)$ with respect to $\omega$. As can be seen in Fig.~\ref{fig:D}, the traditional order of convergence is approximately $3.2$.

Given that $e_1$ vanishes in the continuum limit we can set $f_\infty=0$ to find the order of convergence to zero. Using Eq. (\ref{eq:convdef}) with the two highest resolutions $i=\{2,3\}$, and setting $f_\infty=0$, $E$ can be eliminated to give
\begin{equation}\label{eq:conv_to_zero}
\frac{f_3}{f_2} = \frac{\Delta x_3^\omega}{\Delta x_2^\omega}.
\end{equation}
Similarly to before, we can define a deviation factor $\tilde{\mathcal{D}}$,
\begin{equation}\label{eq:zeroconv}
\tilde{\mathcal{D}}(\omega) = \int_{t_0}^{t_1}\left(\frac{f_3}{f_2} - \frac{\Delta x_3^\omega}{\Delta x_2^\omega}\right)^2\,\dd t,
\end{equation}
which time averages the violation of Eq.~(\ref{eq:conv_to_zero}). The black curve in Fig.~\ref{fig:D} plots Eq.~(\ref{eq:zeroconv}) versus $\omega$ and the order of convergence to zero can be estimated by minimising $\tilde{\mathcal{D}}(\omega)$ with respect to $\omega$. As can be seen in Fig.~\ref{fig:D}, the order of convergence convergence to zero is approximately $1.9$.

\section{Conclusion} \label{sect:conclusion}

A derivation of the QFS system (\ref{qfs_system}) for continuity equations, valid locally for general spacetimes, is derived and applied to spherical integration surfaces. Although spherical extraction surfaces are used, the methods of section \ref{sect:sphere} can be applied to general extraction surfaces with minor adjustments. The QFS system is used to calculate the well known Noether charge densities for complex scalar and complex vector (Proca) fields along with novel expressions for the flux variable $\mathcal{F}$ in section \ref{sect:noether}. Next the QFS system for energy momentum currents associated with matter are found and the main result of this paper is the explicit derivation of the angular momentum QFS variables $\mathcal{Q}$,  $\mathcal{F}$ and $\mathcal{S}$. The three variables can be used to measure the angular momentum of matter within a region, the flux of angular momentum of matter through the boundary of that region and the transfer of angular momentum between matter and curvature; they can also be used with Eq.~(\ref{qfs_system}) to determine the numerical quality of a simulation as the QFS system is exactly satisfied in the continuum limit. In section \ref{sect:results} the combination of variables $\mathcal{Q}$ and $\mathcal{S}$ is shown to be a superior measure of angular momentum than integrals of only the charge density $\mathcal{Q}$ in two ways; firstly its measurement is less prone to oscillations and secondly it is conserved in the large radius limit.

The QFS system for angular momentum is numerically tested on a dynamic non-linear spacetime consisting of two colliding boson stars; the collision has a small impact parameter giving rise to a non-zero total angular momentum. The stars promptly collide and form a highly perturbed, localised scalar field configuration partially retaining angular momentum. The total angular momentum of the spacetime is measured using the QFS variables (Eqs.~(\ref{bigq}), (\ref{final_flux}) and (\ref{s_explicit_angmom})) and is shown to agree well with the Newtonian approximation. This is a good check on the normalisation of the QFS variables as they should return the Newtonian calculation in the low energy limit; even though we simulate a fully non-linear spacetime the density and boost velocity of the stars are mild. The final numerical result is the convergence test of the QFS system which measures the relative error described in \ref{sect:results}. The relative error converges to zero with order $\omega\approx 1.9$ in the continuum limit and the highest resolution simulation gives a fractional error of approximately $3 \%$ in the total angular momentum after $8000$ time units. 

The QFS system is straightforward to implement and it is hoped these results will be useful to the Numerical Relativity community for better measurement of local energy-momentum of matter and Noether charge aswell as powerful check on simulation resolution.

\section*{Acknowledgements} \label{sect:thanks}
I would like to thank Katy Clough, Bo-Xuan Ge, Thomas Helfer, Eugene Lim, Miren Radia and Ulrich Sperhake for many helpful conversations. This work is supported by STFC-CDT PhD funding, PRACE Grant No. 2020225359 and DIRAC RAC13 Grant
No. ACTP238. Computations were performed on the Cambridge Service for Data Driven Discovery (CSD3) system, the Data Intensive at Leicester (DIaL3) and the Juwels cluster at GCS@FZJ, Germany.

\appendix

\section{Source Term Calculation} \label{sect:source}
Here we expand the source term $\mathcal{S}$ from section \ref{sect:derive},
\begin{align}
\mathcal{S} &= \alpha T_{\mu\nu} \nabla^\mu \xi^\nu.
\end{align}
Note that $\bs{\xi}$ is assumed spacelike, $\xi^\mu n_\mu = 0\rightarrow \xi^0=0$. If the reader is interested in a timelike $\bs{\xi}$, for calculating the source term of energy, it can be found in \cite{Clough_2021}. Expanding the stress tensor with the usual 3+1 components \cite{gourgoulhon20073+}, \cite{alcubierre2008introduction} $(S_{\mu\nu}, S_{\mu}, \rho)$, gives
\begin{align}
\frac{1}{\alpha}\mathcal{S} &= S_{\mu\nu}\nabla^\mu \xi^\nu + S_\mu n_\nu \nabla^\mu \xi^\nu + S_\nu n_\mu \nabla^\mu \xi^\nu + \rho n_\mu n_\nu\nabla^\mu \xi^\nu.
\end{align}
Let us decompose each piece separately. Starting with the spacelike tensor $S_{\mu\nu}$ term,
\begin{align}
S_{\mu\nu}\nabla^\mu \xi^\nu &= (S_{\rho \sigma}\perp^\rho _\mu \perp^\sigma_\nu )\nabla^\mu ( \perp^\nu_n \xi^n ),  \\
&= S_{\rho \sigma}(\perp^\rho _\mu \perp^b_\nu \nabla^\mu ( \perp^\nu_n \xi^n )),  \\
&= S_{\mu\nu} D^\mu \xi^\nu, \\
&= S^i_{j}\partial_i \xi^j + S^i_{j} {}^{(3)}\Gamma^j_{\,\,\,i k} \xi^k, 
\label{eq:S1}
\end{align}
where we used the idempotence of the projector $\bs{\perp}$ on components $S_{\mu\nu}$ and $\xi^\mu$ which are already projected onto $\Sigma$. Here $\bs{D}$ and ${}^{(3)}\Gamma^j_{\,\,\,i k}$ are the covariant derivative and Christoffel symbol components of $\Sigma$. Some algebra shows that the terms containing $S_\mu$ become
\begin{align} 
\label{eq:S2}S_\nu n^\mu \nabla_\mu \xi^\nu + S^\mu n_\nu \nabla_\mu \xi^\nu  &= S_\nu \mathcal{L}_n \xi^\nu, 
\end{align}
where we used the fact that $S^0 = 0$, $n_{i\neq0}=0$ and that the we are free to swap between $\partial_\mu \leftrightarrow \nabla_\mu$ derivatives in a Lie derivative. Finally the $\rho$ term simplifies, using $\nabla_\mu (n^\nu n_\nu) = 0$, to
\begin{align} \label{eq:S3}
\rho n_\mu n_\nu\nabla^\mu \xi^\nu &= \rho n_\nu \mathcal{L}_n \xi^\nu.
\end{align}
Combining Eqs.~(\ref{eq:S1}), (\ref{eq:S2}) and (\ref{eq:S3}) we can write the source term as,
\begin{align}
\frac{1}{\alpha}\mathcal{S} &= S^i_{j}\partial_i \xi^j + S^i_{j} {}^{(3)}\Gamma^j_{\,\,\,i k} \xi^k + S_\nu \mathcal{L}_n \xi^\nu + \rho n_\nu \mathcal{L}_n \xi^\nu,
\end{align}
We can expand the Lie derivatives to partial derivatives, for ease of numerical implementation, with the following assumptions $n_\mu S^\mu = 0$, $\xi^0 = 0$, $n_{i\neq0}=0$ and $\partial \xi^0 = 0$.
\begin{align}
S_\nu \mathcal{L}_n \xi^\nu &= -\frac{1}{\alpha} S_\nu \beta^i \partial_i \xi^\nu  + \frac{1}{\alpha}S_\nu \xi^\mu \partial_\mu \beta^\nu \\
\rho n_\nu \mathcal{L}_n \xi^\nu &= -\frac{1}{\alpha} \rho \xi^\mu \partial_\mu \alpha 
\end{align}
This gives us our final form for the angular momentum source density,
\begin{align}\label{s_explicit} 
\begin{split}
\mathcal{S} &= \alpha S^\mu_{\nu}{}^{(3)}\partial_\mu \xi^\nu + \alpha S^\mu_{\nu} {}^{(3)}\Gamma^\nu_{\,\,\,\mu \sigma} \xi^\sigma \\&\quad- S_\nu \beta^i \partial_i \xi^\nu  + S_\nu \xi^\mu \partial_\mu \beta^\nu - \rho \xi^\mu \partial_\mu \alpha.
\end{split}
\end{align}
If we pick a coordinate basis vector as our approximate Killing vector, for example with components $\xi^\mu = (\partial_\phi)^\mu =(0,0,0,1)^\mu$ in polar coordinates, then the $\partial_\mu \xi^\nu$ terms will vanish. However if we wish to work in Cartesian coordinates, which is very common for numerical codes, then the vector components $\tilde\xi^\mu$ become,
\begin{align}
\tilde \xi ^\mu  &= (\partial_\phi)^\nu \frac{\partial \tilde x^\mu}{\partial x^\nu}=  (0,-y,x,0),
\end{align}
where $\tilde x^\mu$ are Cartesian coordinates and $x^\mu$ are spherical polar coordinates.

\section{Generality of Result} \label{sect:generality}
\begin{figure}[h]
{\includegraphics[width=1.0\columnwidth]{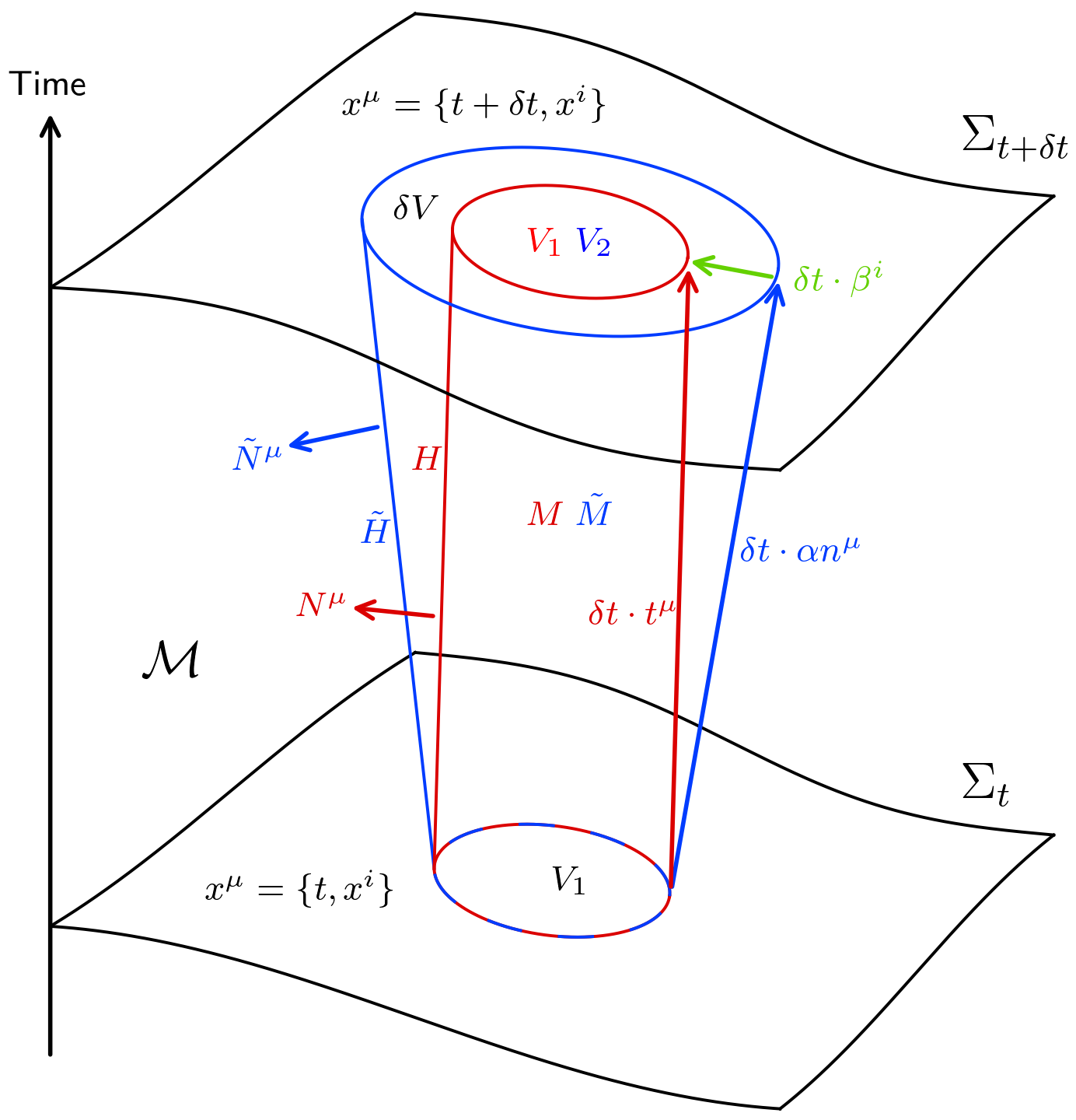}}
\caption{Comparison of two possible geometries for derivation of QFS system in section \ref{sect:derive} on manifold $\mathcal{M}$. $\Sigma_t$ is the spatial hypersurface at time $t$ and $\Sigma_{t+\delta t}$ is the the spatial hypersurface at a later time $t+\delta t$. $V_1$ is the coordinate volume, with surface $\partial V_1$ (not labelled), that we wish to use as an extraction volume on $\Sigma_t$. The red cylinder, defined by $\partial V_1$ evolved along integral curves of $\bs{t}=\bs{\partial_t}$, is the same as in Fig.~\ref{fig:qfs_geometry}. Evolving $\partial V_1$ forward in time with $\bs n$, as demonstrated with the blue cylinder, gives a different coordinate volume $\partial V_2$ (not labelled) on $\Sigma_{t+\delta t}$. Similarly to the red cylinder, the blue cylinder has sides labelled by $\tilde{H}$ and an interior $\tilde{M}$. The difference in the volumes $V_1$ and $V_2$ on $\Sigma_{t+\delta t}$ is denoted by $\delta V.$ }
\label{fig:qfs_comparison}
\end{figure}

Here we demonstrate that the choice of 4-volume $M$ integrated in Eq.~(\ref{master_integral}) does not change the resulting QFS system (\ref{first_qfs}). We start by defining the extraction 3-volume $V_1\in\Sigma$ at time $t=t_0$. The boundary of $V_1$ is the 2-volume $\partial V_1$ with metric $\bs{\sigma}$. As in section \ref{sect:derive}, $\Sigma$ is the 3-manifold defined by the set of all points with constant time coordinate $t$, equipped with metric $\bs{\gamma}$ and unit normal $\bs{n}$ like Eq.~(\ref{n_def}). We now choose to define a 4-volume $\tilde{M}$, different to $M$, as the evolution of $V_1$ along integral curves of $\bs{n}$ between times $t_0 \leq t \leq t_0 + \delta t$ in the limit $\delta t \rightarrow 0$. The boundary of $\tilde{M}$, $\partial \tilde M$, is composed of three coordinate 3-volumes, $V_1\in\Sigma_t$, $V_2\in\Sigma_{t+\delta t}$ and $\tilde{H}$. Here $V_2=V_1 + \delta V$ and is the future of $V_1$, at time $t=t_0 +\delta t$, found by following integral curves of $\bs{n}$. $\tilde{H}$ is the 3-volume defined by the time evolution of 2-volume $\partial V_1$ with $\bs{n}$. A diagram showing the differences between the choices of time evolution vectors $\bs{n}$ and $\bs{t}$ is given in Fig.~\ref{fig:qfs_comparison}.

We start again by using Gauss' theorem like in Eq.~(\ref{master_gauss}) which results in three surface integrals over $V_1\in\Sigma_t$, $V_2\in\Sigma_{t+\delta t}$, and $\tilde{H}$,
\begin{align} \begin{split}
\int_{\tilde{M}} \bs{\nabla} \cdot \bs{J} \sqrt{-g} \, \dd^4 x =& -\int^{t=t_0+\delta t}_{V_2} \bs{n} \cdot \bs{J} \sqrt{\gamma} \,\dd^3 x \\
                                                             & +\int^{t=t_0}_{V_1} \bs{n} \cdot \bs{J} \sqrt{\gamma} \,\dd^3 x \\ 
                                                             & +\int_{\tilde{H}} \tilde{\bs N} \cdot \bs{J} \sqrt{-\tilde{h}} \, \dd x^2 \dd t,
\end{split}\end{align}
where $\bs{\tilde{N}}$ is the unit normal to $\tilde{H}$. Starting with the integrals over $V_1$ and $V_2 = V_1 + \delta V$ we get, 
\begin{align}
-\int^{t=t_0+\delta t}_{V_2} \bs{n} \cdot \bs{J} \sqrt{\gamma} \,\dd^3 x  &+\int^{t=t_0}_{V_1} \bs{n} \cdot \bs{J} \sqrt{\gamma} \,\dd^3 x, \\ 
 \label{dsigma dt bit}=-\delta t \,\partial_t \int_{V_1} \bs{n} \cdot \bs{J} \sqrt{\gamma} \,\dd^3 x  &-\int_{\delta V} \bs{n} \cdot \bs{J} \sqrt{\gamma} \,\dd^3 x, 
\end{align}
with the new integral over $\delta V$ appearing because $V_1\neq V_2$ as $V_1$ is evolved along integral curves of $\bs{n}$ rather than time basis vector $\bs{t}=\bs{\partial_t}$; this is demonstrated in Fig.~\ref{fig:qfs_comparison}. In the limit that $\delta t \rightarrow 0$ it can be seen that,
\begin{align} \label{deltasigmabit}
\int_{\delta V} \bs{n} \cdot \bs{J} \sqrt{\gamma} \,\dd^3 x = -\delta t \int_{\partial V_1} \beta^i s_i\, \bs{n} \cdot \bs{J} \sqrt{\gamma} \,\dd^2 x,
\end{align}
where $s_i$ are the components of the unit normal to the coordinate surface $\partial V_1$ in $\mathbb{R}^3$ rather than $\Sigma_t$. The overall negative sign in Eq.~(\ref{deltasigmabit}) comes from the defined direction of the shift vector $\bs{\beta}$ as seen Fig.~\ref{fig:qfs_comparison}. Addressing the integral over $\tilde{H}$ gives,  
\begin{align}
 \int_{\partial V_1}\int_{t_0}^{t_0 + \delta t}\tilde{\bs N} &\cdot \bs{J} \sqrt{-\tilde{h}} \, \dd x^2 \dd t, \nonumber \\
 \label{htildepart}=\delta t \int_{\partial V_1} \tilde{\bs N} &\cdot \bs{J} \sqrt{-\tilde{h}} \, \dd x^2.  
\end{align}
Combining Eqs.~(\ref{dsigma dt bit}), (\ref{deltasigmabit}) and (\ref{htildepart}) and a source term like in (\ref{master_integral}) we get,
\begin{align}
 \partial_t &\int_{V_1} \bs{n} \cdot \bs{J} \sqrt{\gamma} \,\dd^3 x =\nonumber \\
 &\quad\quad\int_{\partial V_1} \left( \beta^i s_i\, \bs{n} \cdot \bs{J} \sqrt{\gamma}  + \tilde{\bs N} \cdot \bs{J} \sqrt{-\tilde{h}} \right) \, \dd x^2 \nonumber \\
 &\quad-\int_{V_1} S \alpha \sqrt{\gamma} \, \dd^3 x,
\end{align}
which is in the same form as Eq.~(\ref{qfs_system}) with definitions,
\begin{align}
\mathcal{Q} :&= J^\mu n_\mu, \\
\label{unfinished_flux}\mathcal{F} :&= \frac{\sqrt{-\tilde h}}{\sqrt{\sigma}}J^\mu \tilde N_\mu +  \frac{\sqrt{\gamma}}{\sqrt{\sigma}}\beta^i s_i\, \mathcal{Q} , \\
\mathcal{S} :&= \alpha S,
\end{align} 
where we used $\bs{n} \cdot \bs{J} = \mathcal{Q}$ for the flux term. The density term $\mathcal{Q}$ and source term $\mathcal{S}$ are agnostic to our choice of extraction surface and its time evolution so have turned out the same as Eqs.~(\ref{q_def}) and (\ref{source_def}). At first glance the flux term $\mathcal{F}$ seems different to Eq.~(\ref{3plus1f}) but evaluating this term in a coordinate basis will show otherwise.

Choosing a spherical extraction surface as in Section \ref{sect:sphere} and using spherical polar coordinates, $x^\mu = \{t,r,\theta,\phi\}$, $V_1$ becomes the coordinate 3-volume $r\leq r_0$, $t=t_0$. The unit normal $\bs{\tilde{N}}$ satisfies $\bs{\tilde{N}}\cdot \bs{n} =0$ so $\tilde N^\mu = (0,\tilde N^i)$ where, 
\begin{align}
\tilde N_i &= \frac{\nabla_i (r-r_0)}{\sqrt{\gamma^{jk}\nabla_j (r-r_0) \nabla_k (r-r_0)}}, \\ 
 &= (\frac{1}{\sqrt{\gamma^{rr}}},0,0),
\end{align}
and the flat space normal has components is $s^i = s_i = (1,0,0)$ with respect to spherical polar coordinates over a different flat manifold. Using Eqs.~(\ref{sigma metric}) and (\ref{sigma det}) with Cramer's rule for matrix inverse, it can be shown that,
\begin{align}
\gamma^{rr} &= \frac{\det \sigma_{ab}}{ \det \gamma_{ij}} = \frac{\sqrt{\sigma}^2}{\sqrt{\gamma}^2}, \\ 
\label{htt}\tilde{h}^{tt} &= \frac{\det \sigma_{ab}}{\det \tilde h_{ij}} = -\frac{\sqrt{\sigma}^2}{\sqrt{-\tilde h}^2},
\end{align}
where it should be noted that $\tilde h < 0$ and $\sigma > 0$. Deriving Eq.~(\ref{htt}) uses the fact that $\tilde H$ intersects $\Sigma$ on $\partial V_1$ and therefore must have the same line element for variations in angular coordinates; hence $g_{\theta\theta}=\tilde{h}_{\theta\theta}$, $g_{\theta\phi}=\tilde{h}_{\theta\phi}$ and $g_{\phi\phi}=\tilde{h}_{\phi\phi}$. The final component we need is to calculate $\tilde h^{tt}$ which can be done by projecting the 4-metric $\bs g$ onto $\tilde H$ as, 
\begin{align}
{}^{(4)}\tilde{h}^{\mu\nu} &= g^{\mu\nu} - \tilde{N}^\mu \tilde{N}^\nu,\\
{}^{(4)}\tilde{h}^{tt} &= g^{tt} - \tilde{N}^t\tilde{N}^t,\\
         \label{htt_lapse}      &= -\alpha^{-2},
\end{align}
where ${}^{(4)}\tilde{\bs{h}}$ is a 4-tensor belonging to $\mathcal{M}$ and $\tilde{N}^t=0$. Using the pushforward of $\tilde{\bs h}$ on $\tilde H$ to ${}^{(4)}\tilde{\bs h}$ on $\mathcal{M}\vert_{p\in{\tilde H}}$, similarly to Sec.~\ref{sect:sphere}, it can be shown that ${}^{(4)}h^{tt}=h^{tt}$. Equations ~(\ref{htt}) and (\ref{htt_lapse}) combine to give,
\begin{align}
  \sqrt{-\tilde{h}} &= \alpha \sqrt{\sigma},
\end{align}
again noting $\tilde h<0$. Now we can re-write the flux (\ref{unfinished_flux}) term as, 
\begin{align}
\mathcal{F} &= \alpha J_\mu \tilde N^\mu +  \frac{1}{\sqrt{\gamma^{rr}}}\beta^i s_i\, \mathcal{Q} , \\
\mathcal{F} &= \alpha\gamma^{ r\nu} J_\nu \tilde N_r +  \frac{1}{\sqrt{\gamma^{rr}}}\beta^r \, \mathcal{Q} , \\
            &= \frac{1}{\sqrt{\gamma^{rr}}}( \alpha \gamma^{ r\nu} J_\nu +  \beta^r \, \mathcal{Q} ), 
\end{align} 
and this is identical to Eq.~(\ref{3plus1fluxspherical}) found earlier.

\bibliographystyle{ieeetr}
\bibliography{refs.bib}

\end{document}